\documentclass[]{aa}
\usepackage{natbib}         % pour A&A
\usepackage{graphicx}
\usepackage[switch]{lineno} % switch=Put the line into the inner margin
\linenumbers\modulolinenumbers[5]
\bibpunct{(}{)}{;}{a}{}{,} % to follow the A&A style

\newcommand{\nombreglobal}{265}
\newcommand{\nombrefit}{102}
\newcommand{\nombrecompar}{54}
\newcommand{\nombrefitnew}{48} % = fit - compar
\newcommand{\nombretotal}{313} % = global + fit - compar
\newcommand{\nombreapprox}{300} % approximation du nombre total
\newcommand{\ind}[1]{_{\mathrm{#1}}}
\newcommand{\diff}{\mathrm{d}}
\newcommand{\refeq}[1]{(\ref{#1})}

\def\Kepler{\emph{Kepler}}

\def\K{\mathcal{K}}

\def\numax{\nu\ind{max}}
\def\nmax{n\ind{max}}
\def\dnurot{\delta\nu\ind{rot}}
\def\dnurotnl{\delta\nu\ind{rot}{}_{,n,\ell}}

\newcommand{\nm}{n\ind{m}}
\newcommand{\np}{n\ind{p}}
\def\ng{n\ind{g}}

\newcommand\Tg{\Delta\Pi_1}
\def\Dnu{\Delta\nu}
\def\d01{d_{01}}

\def\coeff{\eta}

\def\gmmode{g-m mode}\def\pmmode{p-m mode}

\def\xir{\xi\ind{r}}\def\xih{\xi\ind{h}}

\def\Frot{\mathcal{R}}
\def\minir{\lambda}

\newcommand{\OmegaK}{\langle\Omega\ind{K}\rangle}
\newcommand{\Omegamoy}{\langle\Omega\rangle}
\newcommand\Trotcore{\langle T\ind{rot}\rangle\ind{c}}

\def\dnusplit{\delta\nu\ind{split}}
\def\nl{_{n,\ell}}
\def\env{\ind{env}}
\def\core{\ind{core}}
\def\xcore{x\core}
\def\envnl{_{\mathrm{env,\,}n,\ell}}
\def\corenl{_{\mathrm{core,\,}n,\ell}}
\def\coefrotg{\alpha\ind{rot}}

\def\rotc{\delta\nu\ind{c}}
\def\rots{\delta\nu\ind{s}}
\def\rotg{\delta\nu\ind{g}}
\def\rotp{\delta\nu\ind{p}}
\def\xp{x\ind{p}}
\def\xg{x\ind{g}}
\def\gg{\gamma\ind{g}} \def\gp{\gamma\ind{p}}

\begin{document}
\title{Spin down of the core rotation in red giants}
\titlerunning{Rotation in red giants}
\author{B. Mosser\inst{1}\and
M.J. Goupil\inst{1} \and
K. Belkacem\inst{1} \and
J.P. Marques\inst{2}\and
P.G. Beck\inst{3} \and
S. Bloemen\inst{3} \and
J. De Ridder\inst{3}\and
C. Barban\inst{1}\and
S. Deheuvels\inst{4}\and
Y. Elsworth\inst{5}\and
S. Hekker\inst{6,5}\and
T. Kallinger\inst{3}\and
R.M. Ouazzani\inst{7,1}\and
M. Pinsonneault\inst{8}\and
R. Samadi\inst{1}\and
D. Stello\inst{9}\and
R.A. Garc{\'\i}a\inst{10}\and
T.C. Klaus\inst{11}\and
J. Li\inst{12}\and
S. Mathur\inst{13}\and
R.L. Morris\inst{12}
}
\offprints{B. Mosser}

\institute{LESIA, CNRS, Universit\'e Pierre et Marie Curie, Universit\'e Denis Diderot,
Observatoire de Paris, 92195 Meudon cedex, France; \email{benoit.mosser@obspm.fr}
\and Georg-August-Universit\"at G\"{o}ttingen, Institut f\"ur Astrophysik, Friedrich-Hund-Platz 1, D-37077 G\"{o}ttingen, Germany
\and Instituut voor Sterrenkunde, K. U. Leuven, Celestijnenlaan 200D, 3001 Leuven, Belgium
\and Department of Astronomy, Yale University, P.O. Box 208101, New Haven, CT 06520-8101, USA
\and School of Physics and Astronomy, University of Birmingham, Edgbaston, Birmingham B15 2TT, United Kingdom
\and Astronomical Institute `Anton Pannekoek', University of Amsterdam, Science Park 904, 1098 XH Amsterdam, The Netherlands
\and Institut d'Astrophysique et de G\'eophysique de l'Universit\'e de Li\`ege, All\'ee du 6 Ao\^ut 17, 4000 Li\`ege, Belgium
\and Department of Astronomy, The Ohio State University, Columbus, OH 43210, USA
\and Sydney Institute for Astronomy, School of Physics, University of Sydney, NSW 2006, Australia
\and Laboratoire AIM, CEA/DSM – CNRS - Universit\'e Denis Diderot – IRFU/SAp, 91191 Gif-sur-Yvette Cedex, France
\and Orbital Sciences Corporation/NASA Ames Research Center, Moffett Field, CA 94035, USA
\and SETI Institute/NASA Ames Research Center, Moffett Field, CA 94035, USA
\and High Altitude Observatory, NCAR, P.O. Box 3000, Boulder, CO 80307, USA
}

%\date{Submitted to A\&A}

\abstract{The space mission \Kepler\ provides us with long and
uninterrupted photometric time series of red giants. We are now
able to probe the rotational behaviour in their deep interiors
using the observations of mixed modes.}
{We aim to measure the rotational splittings in red giants and to
derive scaling relations for rotation related to seismic and
fundamental stellar parameters.}
{We have developed a dedicated method for automated measurements
of the rotational splittings in a large number of red giants.
Ensemble asteroseismology, namely the examination of a large
number of red giants at different stages of their evolution,
allows us to derive global information on stellar evolution.
}%
{We have measured rotational splittings in a sample of about
\nombreapprox\ red giants. We have also shown that these
splittings are dominated by the core rotation. Under the
assumption that a linear analysis can provide the rotational
splitting, we observe a small increase of the core rotation of
stars ascending the red giant branch. Alternatively, an important
slow down is observed for red-clump stars compared to the red
giant branch. We also show that, at fixed stellar radius, the
specific angular momentum increases with increasing stellar mass.}
{Ensemble asteroseismology indicates what has been indirectly
suspected for a while: our interpretation of the observed
rotational splittings leads to the conclusion that the mean core
rotation significantly slows down during the red giant phase. The
slow-down occurs in the last stages of the red giant branch. This
spinning down explains, for instance, the long rotation periods
measured in white dwarfs.}

\keywords{Stars: oscillations - Stars: interiors - Stars:
rotation - Stars: late-type}
%- Methods: data analysis - Methods: analytical}

\maketitle\voffset = 1.2cm
%________________________________________________________________

\section{Introduction\label{introduction}}

The internal structure of red giants bears the history of their
evolution. They are therefore seen as key for the understanding of
stellar evolution. They are expected to have a rapidly rotating
core and a slowly rotating envelope
\citep[e.g.][]{2000ApJ...540..489S}, as a result of internal
angular momentum distribution. Indirect indications of the
internal angular momentum are given by surface-abundance anomalies
resulting from the action of internal transport processes and from
the redistribution of angular momentum and chemical elements
\citep{1992A&A...265..115Z,2008A&A...482..597T,2009pfer.book.....M,2011A&A...527A..94C}.
Direct measurements of the surface rotation are given by the
measure of $v\,\sin i$ \citep[e.g.][]{2008AJ....135..892C}. The
slow rotation rate in low-mass white dwarfs
\citep[e.g.][]{1999ApJ...516..349K} suggests a spinning down of
the rotation during the red giant branch (RGB) phase. In addition,
3-D simulations show non-rigid rotation in the convective envelope
of red giants \citep{2009ApJ...702.1078B}. Different mechanisms
for spinning down the core have been investigated
\citep[e.g.][]{2005Sci...309.2189C}. Rotationally-induced mixing,
amid other angular momentum transport mechanisms, is still poorly
understood but is known to take place in stellar interiors.
Therefore, a direct measurement of rotation inside red giants
would give us an unprecedented opportunity to perform a leap
forward on our understanding of angular momentum transport in
stellar interiors
%\citep[e.g.][]{2010A&A...509A..72E,2012A&A...543A.108L,2012A&A...544L...4E}.
\citep[e.g.][]{2012A&A...543A.108L,2012A&A...544L...4E}.

\begin{figure}
%multrot, f, p, 6144777, 7
\includegraphics[width=8.8cm]{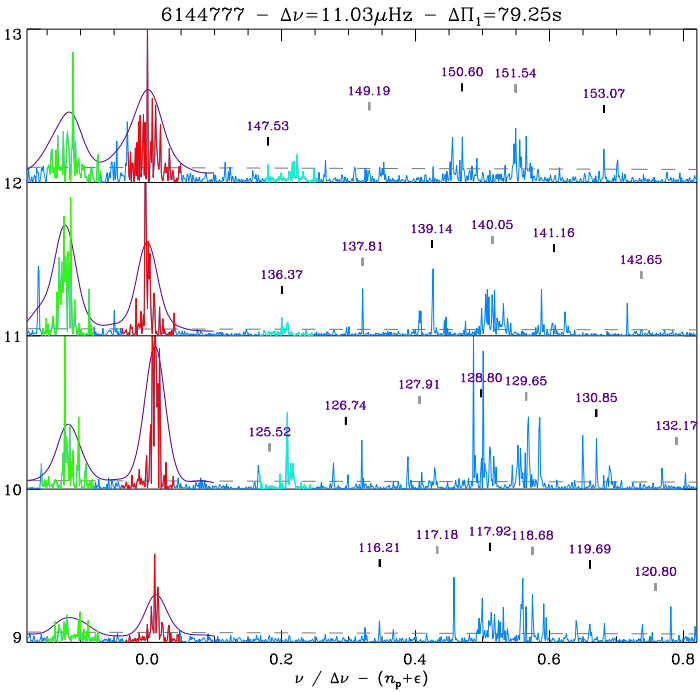}
\caption{\'Echelle diagram of a typical RGB star (KIC 6144777) as
a function of $\nu/\Dnu - (\np+\varepsilon)$. The radial order
$\np$ is indicated on the y-axis. Radial modes (highlighted in
red) are centered on 0, quadrupole modes (highlighted in green),
near $-0.12$ (with a radial order $\np-1$), and $\ell=3$ modes,
sometimes observed, (highlighted in light blue) near 0.20. Dipole
mixed modes are identified with the frequency given by the
asymptotic relation of mixed modes, in $\mu$Hz. The fit is based
on  peaks showing a height larger than eight times the mean
background value (grey dashed lines).
 \label{figeps}}
\end{figure}

This is becoming possible with seismology, which provides us with
direct access to measure the internal rotation profile, as shown
by \cite{2012Natur.481...55B} and \cite{2012ApJ...756...19D}. They
demonstrate that it is possible to  measure the core rotation
thanks to the precision of asteroseismic results derived from the
photometric light curves of red giants provided by the NASA
\Kepler\ mission \citep{2010Sci...327..977B}. Previous work on
field star asteroseismic surveys shows that is possible to define
evolutionary sequences \citep[e.g.][]{2009A&A...503L..21M,
2011ApJ...743..143H}. So, with ensemble asteroseismology, we aim
to get measurements in a large enough sample of red giants to
explore the variation of the internal rotation with evolution.

After a decade of relatively uncertain ground-based measurements,
CoRoT has unambiguously revealed that red giants show solar-like
oscillations \citep{2009Natur.459..398D}. Important scaling
relations have then been shown \citep{2009A&A...506..465H} for
deriving crucial information on the stellar mass and radius from
global seismic parameters \citep{2010A&A...522A...1K}. Specific
features of solar-like oscillations in red giants have been
characterized
\citep[e.g.][]{2010ApJ...713L.176B,2010A&A...517A..22M,2010ApJ...723.1607H}.
Mixed modes, which correspond to the coupling of gravity waves in
the radiative core region and pressure waves in the envelope
\citep{2001MNRAS.328..601D,2009A&A...506...57D}, have been
detected in red giants. They were first reported by
\cite{2010ApJ...713L.176B}. Their period spacings were measured by
\cite{2011Sci...332..205B}. \cite{2011Natur.471..608B} and Mosser
et al. (2011a) have shown the capability of these modes to measure
the evolutionary status of red giants. Mixed modes can be divided
into two categories, namely gravity-dominated mixed modes
(hereafter called \gmmode s) which have large amplitudes in the
core, and, in contrast, pressure-dominated mixed modes (\pmmode
s). The frequencies of the \pmmode s are very close to the
theoretical pure p mode frequencies; they however appear to have a
significant g component, and are therefore sensitive also to the
core conditions. We analyse in this work mostly stars showing a
rich mixed-mode spectrum.

Observations and data are presented in Section \ref{data}. The
observed properties of the rotational splittings are described in
Section \ref{analysis}. In Section \ref{scaling}, we derive
scaling relations governing the rotational splitting, independent
of any modeling, but based on the observational evidence of a much
higher rotation rate in the stellar core. The way the core
rotation is related to the measured rotational splitting is
quantified in Section \ref{splitting-rotation}. In Section
\ref{discussion}, we then derive unique information on the core
rotation in red giants and probe their internal angular momentum
and its evolution.

%===================================================================
\section{Data\label{data}}

\subsection{25-month long observation}

The red giant stars analyzed in this work have already been
presented \citep[e.g.][and references
therein]{2011A&A...525A.131H}. We now benefit from longer time
series. All red giants observed up to \Kepler 's quarter Q10 have
been analyzed. Original light curves were processed according to
\cite{2010ApJ...713L..87J} and corrected according to the
procedure of \cite{2011MNRAS.414L...6G}. The Fourier analysis of
the 868-day long time series provides a frequency resolution of
about 11.5\,nHz. Due to the characteristics of the measurement, we
have in principle access to rotation periods up to the observation
duration. For a few RGB stars showing large rotational splittings,
a reprocessed set of Q0 to Q11 {\it Kepler} data has been used,
based on the extraction of the stellar fluxes using new custom
masks from the recently released pixel-data information (Bloemen
et al. 2012). These new light curves were then corrected applying
the algorithms developed by \cite{2011MNRAS.414L...6G} but using a
refined new automatic procedure (Mathur et al., in preparation).
In practice, we measure rotational splittings with periods in the
range 8 -- 280 days.

\subsection{Mixed-mode pattern}

The complete identification of the red giant pressure oscillation
pattern is given by the description of the so-called universal
oscillation pattern \citep{2011A&A...525L...9M}. This method
alleviates any problem of mode identification. The whole frequency
pattern of pure p modes is approximated by:
\begin{equation}\label{tassoulp}
\nu_{\np,\ell} = \left(\np +{\ell \over 2}+\varepsilon - d_{0\ell}
 + {\alpha \over 2}\; [ \np- \nmax ]^2 \right) \Dnu,
\end{equation}
where $\Dnu$ is the mean large separation measured in a broad
frequency range around the frequency $\numax$ of maximum power,
$\np$ is the p-mode radial order, $\ell$ is the angular degree,
$\varepsilon$ is the phase offset, $d_{0\ell}$ accounts for the
so-called small separation, $\alpha$ is a small constant, and
$\nmax= \numax / \Dnu$. The parameters $\varepsilon$, $d_{0\ell}$
and $\alpha$ are considered as a function of the large separation;
an updated fit of $\varepsilon$ is used, comparable to the
expression of \cite{2012arXiv1205.4023C}. The parameter $\alpha$
represents the second-order term of the asymptotic development
\citep{1980ApJS...43..469T} and accounts for the mean curvature of
the radial mode oscillation pattern \citep{mesure}. It was
considered as a constant by \cite{2011A&A...525L...9M}. Here, with
much longer time series and large separations observed up to
20\,$\mu$Hz, we prefer to use the fit
\begin{equation}\label{fit_alpha}
\alpha = 0.015 \ \Dnu^{-0.32},
\end{equation}
with $\Dnu$ in $\mu$Hz. This relation is derived from the detailed
analysis of the radial modes with the method presented by
\cite{2010AN....331..944M}.

The recent analysis of \cite{2012A&A...541A..51K} has shown that
this development is valid, independent of the stellar evolutionary
status, under the condition that the determination of the large
separation is global and not local. Their Figure 6 illustrates the
important curvature of the ridges depicted by the quadratic term
of Eq.~\refeq{tassoulp}. The comparative work by
\cite{2012A&A...544A..90H} shows that the extra hypothesis used by
\cite{2011A&A...525L...9M}, expressed by the function
$\varepsilon( \Dnu)$, helps to obtain very precise results. Using
this constraint requires fitting each oscillation spectrum in a
large frequency range around $\numax$ and taking into account the
mean curvature of the p-mode oscillation pattern.
%This curvature can be seen as the second order term of the asymptotic relation, with
%$\Dnu$ in Eq.~\ref{tassoulp} measured at $\numax$ and therefore not equal to the asymptotic large separation.

Equation \refeq{tassoulp} holds precisely for radial modes and
gives a proxy for the pure pressure dipole ($\ell=1$) modes.
However, due to the significant coupling of pressure waves with
gravity waves in the inner radiative region, the dipole
oscillation pattern is dominated by mixed modes located around the
position of the pure p mode. The frequency interval between the
individual mixed modes is not constant and is determined according
to the method presented in \cite{2011A&A...532A..86M}. For stars
showing a large number of g-m modes, the oscillation pattern can
be precisely described by an asymptotic relation presented by
\cite{goupil}, following the ideas originally developed by
\cite{1979PASJ...31...87S} and \cite{1989nos..book.....U}. We
apply this method, as explained in \cite{2012A&A...540A.143M}, in
order to locate precisely the dipole mixed modes.

The \'echelle diagram of Fig.~\ref{figeps} shows the
identification of the radial modes provided by
Eq.~\refeq{tassoulp}, with the mode curvature, and the location of
the mixed $\ell=1, m=0$ modes defined by the asymptotic relation.
The remaining shifts between the actual and expected peaks
positions, due for instance to a sharp structure variation
\citep{2010A&A...520L...6M}, are small compared to the mixed mode
spacings. Hence, they do not hamper the mode identification.

\begin{figure}
\includegraphics[width=8.8cm]{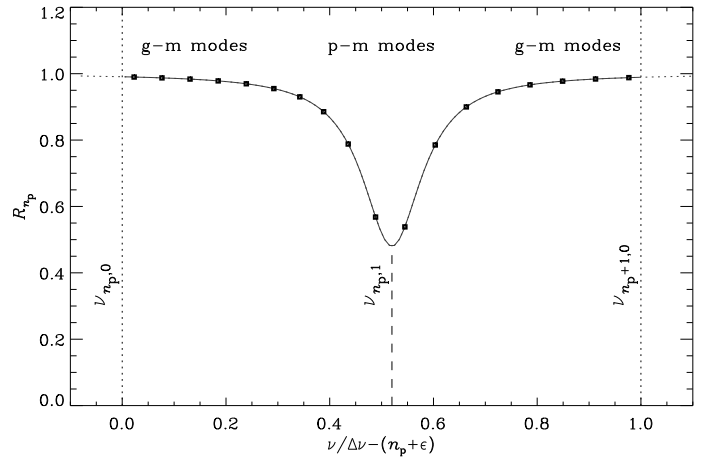}
\caption{Empirical rotation profile $\Frot_{\np} $ for dipole
mixed modes associated to the pure p mode of radial order $\np$,
as a function of the reduced frequency $\nu / \Dnu -
(\np+\varepsilon)$. \label{profil}}
\end{figure}

%\begin{figure*}
%\includegraphics[width=16.88cm]{f7a.ps}
%\includegraphics[width=16.88cm]{f7b.ps}
%\caption{Fit of the rotational multiplets based on the identification of the \gmmode\ pattern and on the modulation of
%the rotational splitting, for the red giants KIC 12008916 and KIC 5858947 \citep[from][]{2012A&A...540A.143M}. The background shaded
%pattern indicates the location of the p modes with different degrees, according to the universal pattern. \label{rot_modele}}
%\end{figure*}

%===================================================================
\section{Measuring the rotational splittings\label{analysis}}

From the analysis of dipole mixed modes,
\cite{2012Natur.481...55B} showed differential rotation in red
giants. In this Section, we propose a model for describing the
rotational splittings of dipole mixed modes and an automated
method for measuring them.

\subsection{Modulation of rotational splittings}

Because of differential rotation, a simple interpretation of
rotational splittings in terms of mean rotation of the star is
inadequate. The situation is made even more complicated by the
fact that rotational splittings are measured for dipole mixed
modes. Indeed, the mixed nature of the modes varies as a function
of mode frequency, as shown by \cite{2009A&A...506...57D}. As a
result of those two interlaced effects, \cite{2012A&A...540A.143M}
have shown that rotational splittings are modulated in frequency,
with a period of $\Dnu$. The rotational splitting can be written
\begin{equation}\label{splitt}
\dnusplit = \nu_{n,1,m} - \nu_{n,1} = m \ \Frot (\nu)\ \dnurot ,
\end{equation}
where $m$ is the azimuthal order and $\dnurot$ is the maximum
value observed for \gmmode s. Locally, around the position of a
pure dipole p mode of radial order $\np$,
\cite{2012A&A...540A.143M} have shown that $\Frot$ can be
empirically expressed as:
\begin{equation}\label{modulation}
 \Frot_{\np} (\nu)
 = 1 - {\minir \over 1 +  {\displaystyle{\left(
 { \nu -   \nu_{\np, 1} \over \beta \Dnu}
   \right)^2}}}
\end{equation}
for a mixed mode with frequency $\nu$ associated with the pressure
radial order $\np$. The development, solely derived from the
observed splittings,  has currently no theoretical basis: the
Lorentzian form has been chosen similar to the observed variation
of the mixed-mode spacing with frequency
\citep{2011Sci...332..205B,2011Natur.471..608B}.

The form of $ \Frot$ (Fig.~\ref{profil}) implies that the
splittings are not symmetric, consistent with the findings of
\cite{deheuvels}. For multiplets near \pmmode s, the closest
component to the theoretical pure p mode has the smallest
splitting. The asymmetry is expressed when considering
Eq.~\refeq{splitt} as an implicit equation, the splitting of the
frequency $\nu_{n,1,m}$ compared to the non-rotating reference
$\nu_{n,1}$ depending on $\nu_{n,1,m}$:
\begin{equation}\label{splitt2}
\nu_{n,1,m} = \nu_{n,1} +  m \ \Frot
(\nu_{n,1,m})\ \dnurot .
\end{equation}
It assumes, as observed but with the limitation of the observed
frequency resolution, that the modulation of the splitting is the
same for all radial orders. In fact, the fits of the splittings,
obtained by supposing that the frequency of the $m=0$ component of
the dipole mixed modes is given by the asymptotic development,
show that the terms $\minir$ and $\beta$ are independent of the
frequency. Furthermore, we have verified that the values of
$\minir$ and $\beta$ are always very close to 0.5 and 0.08,
respectively, and hence do not depend on the evolutionary stage.
%Figure \ref{rot_modele}, from \cite{2012A&A...540A.143M}, shows
%the fit of the multiplets in typical red giants as predicted by
%Eqs.~\refeq{splitt} and \refeq{modulation}.

\begin{figure}
\centering
\includegraphics[width=8.8cm]{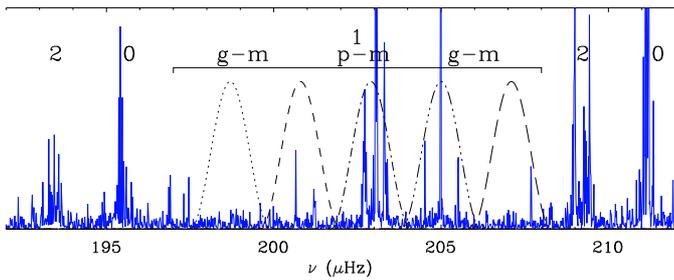}
\caption{Zoom on the oscillation spectrum of the target KIC
10777816. Different narrow filters centered in the $\ell= 1$ mixed
mode range, indicated with different line styles, allow us to
measure a local rotational splitting in each filter. For clarity,
only those filters centered on possible multiplets have been
represented. \label{exemple}}
\end{figure}

Due to the large volume of data, we need an automated method for
deriving the rotational splittings, similar to the determination
of the global seismic parameters $\Dnu$ and $\numax$
\citep[e.g.][]{ 2011A&A...525A.131H,2011MNRAS.415.3539V}. This
method has to cope with an interweaving of rotational splittings
and mixed-mode spacings.

\subsection{EACF analysis}

We can use the envelope autocorrelation function (EACF) method
\citep{2009A&A...508..877M} to measure the rotational splittings
of non-radial modes. This method was developed to measure the
frequency spacing corresponding to the large separation of
solar-like oscillations. With narrow filters centered on the
expected $\ell=1$ pressure modes, it gives the spacing due to the
mixed-mode pattern \citep{2011A&A...532A..86M}. With ultra-narrow
filters centered on each individual mixed mode, it proves to be
able to measure the rotational splittings. The central positions
of the filters are chosen within the frequency ranges where mixed
modes are expected. Each range is wider than half the large
separation $\Dnu$, as shown by \cite{2012A&A...537A..30M}.
Different central positions of the filters are tested within this
range, independent of the mixed-mode positions. The widths of the
filters have to be narrow enough to select only one mixed mode, in
order to avoid confusion between the g-mode spacing and the
rotational splitting (Fig.~\ref{exemple}). They are varied, in
order to test a large range of splittings. We have tested five
different widths, of the order of 1\,$\mu$Hz or less, which
encompass the observed values.

In the EACF method, the signature of any comb-like structure in
the spectrum, here the rotational splitting, is provided by the
highest peak in the spectrum of the windowed spectrum. The signal
is normalized in such a way that the mean white-noise level is 1.
The reliability of the detection is derived from an H0 test. Since
we use ultra-narrow filters, only high signal-to-noise time series
can be analyzed.

For each filter width, multiplets were searched in seven radial
orders around $\numax$, at ten different positions per radial
order. As a consequence, seventy possible signatures of the
rotational splittings are analysed for a given filter. The
criterion for a positive detection is the measurement of similar
splittings for at least three positions centered on different
dipole mixed modes corresponding to different radial orders $\np$,
with a signature of the EACF above the threshold level
\citep{2009A&A...508..877M}. In order to account for the
modulation of $\Frot$ towards \pmmode s (Fig.~\ref{profil}), we
allow relative differences of 1/3 between the individual
measurements. This threshold level has been determined
empirically. It takes into account the fact that splittings are
measured mainly in the wings of the function $\Frot$: the
splittings of \pmmode s are not easily measured, due to their
shorter lifetime, whereas the amplitudes of the \gmmode s located
far from the \pmmode s are too low to give a reliable signature.

\begin{figure}
\includegraphics[width=8.8cm]{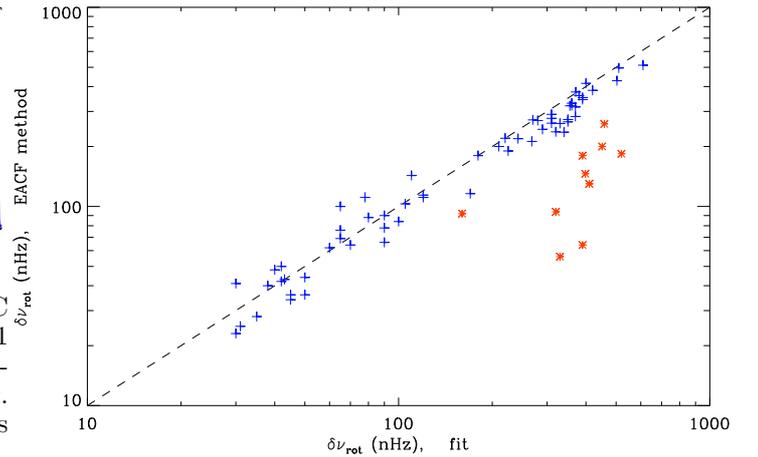}
\caption{Comparison of the splittings measured with the EACF
automated method to the splittings measured with the fit of the
mixed modes (Eq.~\refeq{modulation}). Red asterisks correspond to
rotational splittings too large to be accurately measured in an
automated way and hence excluded from the sample.
 \label{comparefit}}
\end{figure}

\begin{figure}
\includegraphics[width=8.8cm]{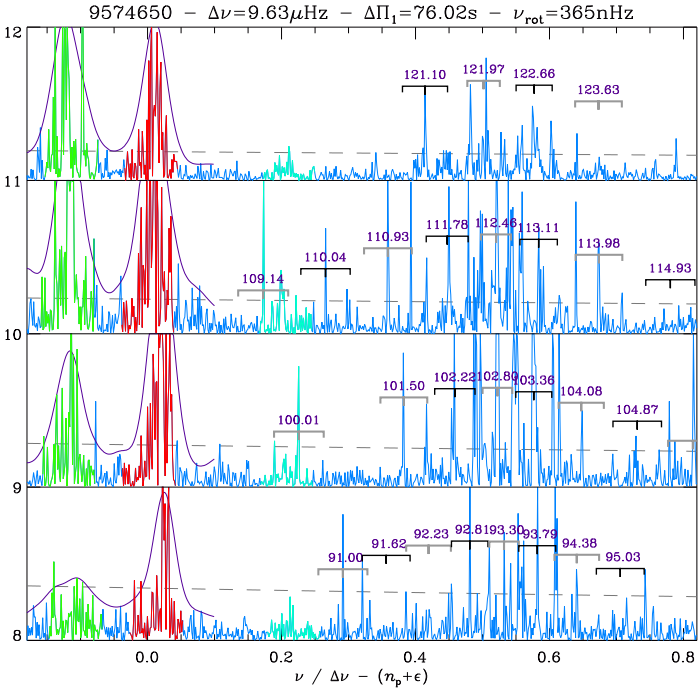}
\caption{Zoom on the rotational splittings of the mixed modes
corresponding to the radial orders $\np = 8 \to 11$ in the giant
KIC 9574650, in an \'echelle diagram as a function of the reduced
frequency $\nu/\Dnu - (\np+\varepsilon)$. All triplets
($m=-1,0,1$) are identified with the function $\Frot$ defined by
Eq.~\refeq{modulation}. At low frequency, multiplets are
overlapping.
 \label{figrot0}}
\end{figure}

\subsection{Performance}

All results found by the automated method have been verified
individually by visually comparing the spectrum with itself after a shift
corresponding to the rotational splitting. The method appeared to
be dominated by \gmmode s since they have narrower widths and are
more numerous than \pmmode s. Hence, they contribute much more
efficiently to the EACF signature, as shown by the careful
examination of numerous individual spectra. Therefore, the method
mainly gives the rotational signature of the core. This is also
clear from the close examination of the extracted splittings.

For stars with a low signal-to-noise ratio oscillation spectrum,
generally faint stars or stars with low $\numax$, we sometimes
obtain spurious results. These results are clearly caused by the
stochastic nature of the oscillation excitation, which
occasionally resembles the complex pattern of multiple peaks shown
by the dipole mode, and were discarded. Finally, depending on the
stellar inclination $i$, the multiplets have two components (when
$\sin i$ is close to 1), three components (intermediate $i$
values), or only one component (low $i$). In this latter case,
measuring the splitting is not possible. The correct
identification of the multiplets has been possible in most of the
cases, allowing us to remove the confusion between $\dnusplit$ and
$2\;\dnusplit$.

Possible confusion between rotational splittings and mixed-mode
spacings has been investigated. Such a confusion is usually
eliminated by limiting the width of the filter. However, RGB stars
may show a rotational splitting very close to the g-mode spacing.
We therefore explored the frequency domain where the solutions are
ambiguous. Ambiguous detections are identified by the fact that,
even if the rotational splitting and the mixed-mode spacing are
both modulated in frequency, with a period $\Dnu$, their
signatures are different. On the one hand, the g-mode frequency
spacing varies as $\nu^2$ since it approximately corresponds to a
regular spacing in period
\citep{2011Natur.471..608B,2012A&A...540A.143M}; on the other
hand, the modulation $\Frot$ is the same all along the spectrum
(Eq.~\refeq{modulation}).

As expected, the automated method fails when the rotational
splittings are larger than half the mixed-mode spacings. This
occurs for giants ascending the RGB, with $\Dnu\le 12\ts\mu$Hz.
For these stars, a dedicated method, presented in the next
paragraph, is needed to disentangle the rotational splittings from
the mixed-mode spacings. We finally measured reliable rotational
splittings in \nombreglobal\ red giants with the automated method,
in the clump and in the early stages of the RGB. The combined
effect of the tiny width of the filter and of the limited
frequency resolution makes the method much more precise at high
frequency than at low frequency. The relative precision is of
about 5\,\% in $\dnurot$ for stars with $\Dnu=15\,\mu$Hz and
25\,\% when $\Dnu=5\,\mu$Hz.

\begin{table}
\caption{Positive detections as a function of
$\Dnu$}\label{detection}
\begin{tabular}{rlcccccr}
 \hline
 \multicolumn{2}{c}{$\Dnu$ range}&\multicolumn{3}{c}{Reference}& \multicolumn{2}{c}{Detections}\\
 \multicolumn{2}{c}{($\mu$Hz)} & $(a)$ & $(b)$ &$(c)$ &  $(d)$ &$(e)$\\
 \hline
 0 & 3 & 323 & 323 &  49 &   0 &   0.0\,\% \\
 3 & 4 & 257 & 250 &  76 &  37 &  48.7\,\% \\
 4 & 5 & 412 & 387 & 167 & 146 &  87.4\,\% \\
 5 & 6 &  97 &  86 &  82 &  21 &  25.6\,\% \\
 6 & 8 &  92 &  81 &  75 &  31 &  41.3\,\% \\
 8 &12 &  68 &  51 &  49 &  48 &  98.0\,\% \\
12 &20 &  50 &  38 &  36 &  30 &  83.3\,\% \\
\hline
  \multicolumn{2}{c}{all} &1299 &1216 & 534 & 313\\
 \hline
\end{tabular}

${(a)}$: total number of red giants in a given frequency range.\\
${(b)}$: same as case $(a)$, but spectra with depressed dipole mixed modes are excluded.\\
${(c)}$: same as case $(b)$, but spectra with an EACF signature
less than 100 are also excluded.\\
${(d)}$: number of positive detections in each frequency range.\\
${(e)}$: percentage of positive detections with respect to
reference $(c)$.
\end{table}

\subsection{Validation with direct measurements\label{method_direct}}

The values of the rotational splitting provided by the EACF method
can be compared to a direct fit of the modulation of the splitting
based on Eqs.~\refeq{modulation} and \refeq{splitt}. A tutorial
for fitting the rotational multiplets is given in Appendix
\ref{methode}; different fits are shown, in order to illustrate
cases with rotational splittings lower, comparable or larger than
the mixed-mode spacings. Using such fits, we measured the
rotational splitting $\dnurot$ in \nombrefit\ red giants. Among
them, \nombrecompar\ had already a rotational splitting measured
with the EACF method. The remaining \nombrefitnew\ stars had
rotational splittings too large compared to the mixed modes
spacings to be measurable with the EACF method. As a result, the
total number of red giant stars with splittings measured with one
or the other method is \nombretotal\ (=\nombreglobal $+$\nombrefit
$-$\nombrecompar). The analysis of the detection and non-detection
as a function of the large separation is summarized in
Table~\ref{detection}. Non-detection at large $\Dnu$ occurs for
RGB stars with depressed mixed modes \citep[identified
by][]{2012A&A...537A..30M} or with very low inclination. In this
latter case, the components $m=\pm1$ are too low to allow the
identification of multiplets. At small $\Dnu$, the non-detection
is explained by the limited frequency resolution, but also by the
poorer quality of the spectra, as expressed by the EACF
coefficient \citep{2009A&A...508..877M}. For $\Dnu$ in the ranges
[4, 5\,$\mu$Hz] and [5, 6\,$\mu$Hz], the success of the detection
depends mainly on the evolutionary status: small rotational
splittings in clump stars with large mixed-mode spacings can be
detected more easily than larger rotational splittings of RGB
stars embedded in narrow spacings. We expect the number of
positive detections to increase with prolonged observations: the
frequency resolution will give access to rotational splittings in
the upper RGB, and the better signal-to-noise ratio will allow the
identification of tiny $m=\pm 1$ components at low stellar
inclination.

We have noted that the automated EACF approach provides values
that are 10\,\% smaller than $\dnurot$ derived from the individual
fits (Fig.~\ref{comparefit}). This small correction is consistent
with the different principles of the methods: the inferred value
of $\dnurot$ from Eqs.~\refeq{splitt} and \refeq{modulation} is
larger than any observed splitting, whereas the EACF method, even
if dominated by \gmmode s, also includes narrower multiplets in
the vicinity of the \pmmode s. This indicates that the 10\,\%
difference can be considered as a bias of the automated method.
For homogeneity, all splittings obtained with the EACF method have
been multiplied by a factor 1.10.

We have tested that the differential-rotation term $\Frot$ of
Eq.~\refeq{modulation} holds for red giants in all stages of their
evolution. Residuals of the fit of the splittings are much smaller
than $\dnurot$. We observe a coefficient $\minir$
(Eq.~\refeq{modulation}) of 0.5$\pm$0.1 for an early RGB star such
as KIC 7341231 studied by \cite{2012ApJ...756...19D}, with $\Dnu =
28.9\ts \mu$Hz, as well as for clump stars with $\Dnu \simeq
4.0\ts \mu$Hz. The parameter $\beta$ is about $0.08\pm0.015$ for
all giants, except a very small number of exceptions, at the
bottom of the RGB. Measuring a modulation profile $\Frot$ almost
independent of the evolution may indicate that its origin obeys to
generic properties, similar for all red giants ascending the red
giant branch (RGB).

For clump stars and early RGB stars, we note that the rotational
splittings are significantly smaller than the mixed-mode spacings.
In these cases, the individual fits confirm the automated
measurement. However, when the rotational splittings are larger
than the mixed-mode spacings, the EACF method was ineffective for
extracting the correct large splitting: it derived wrong values
that result from a combination of the mixed-mode spacing and
rotational splitting. Application of Eq.~\refeq{modulation}
allowed us to provide the correct splittings even with large
values comparable to the mixed-mode spacing. We stress that, in
practice, it seems impossible to disentangle such multiplets from
the mixed-mode pattern without the use of the asymptotic relation
of mixed modes. We were able to investigate complicated cases,
with rotational splittings significantly larger than the
mixed-mode spacings, for RGB stars with $\Dnu$ down to 7\,$\mu$Hz.
Finally, even in the case where multiplets overlap, that is
$\nu_{\nm+1,1,-1} < \nu_{\nm,1,+1}$, with $\nm$ the mixed-mode
order, we do not observe any modification of the profile $\Frot$
(Fig.~\ref{figrot0}). This indicates the absence of avoided
crossings between dipole mixed modes with consecutive mixed-mode
orders $\nm$ and different azimuthal orders $m$. In other words,
this shows the absence of coupling between such modes.

\begin{figure}
\centering
\includegraphics[width=8.8cm]{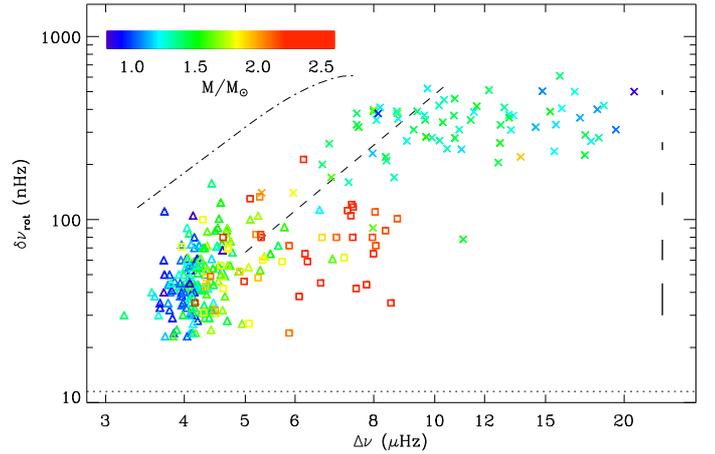}
\caption{Rotational splitting $\dnurot$ as a function of the large
separation $\Dnu$, in log-log scale. The dotted line indicates the
frequency resolution. The dashed and dot-dashed lines represent
the confusion limit with mixed modes in RGB and clump stars,
respectively, derived from \cite{2012A&A...540A.143M}. Crosses
correspond to RGB stars, triangles to clump stars, and squares to
secondary clump stars. The color code gives the mass estimated
from the asteroseismic global parameters. The vertical bars
indicate the mean error bars, as a function of the rotational
splitting. \label{dnu_nurot}}
\end{figure}

\begin{figure}
\includegraphics[width=8.8cm]{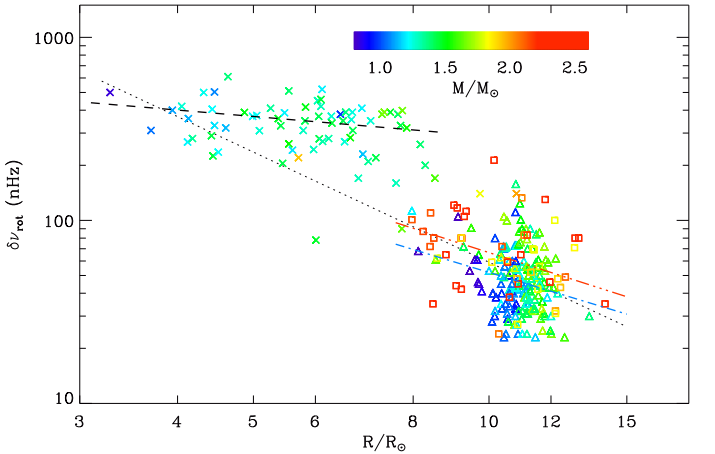}
\caption{Rotation splitting $\dnurot$ as a function of the
asteroseismic stellar radius, in log-log scale. Same symbols and
color code as in Fig.~\ref{dnu_nurot}. The dotted line indicates a
splitting varying as $R^{-2}$. The dashed (dot-dashed,
triple-dot-dashed) line corresponds to the fit of RGB (clump,
secondary clump) splittings.
\label{split_radius_masse}}
\end{figure}

%=======================================================================
\section{Scaling relations\label{scaling}}

We have derived estimates of the stellar mass and radius from the
seismic global parameters and from the effective temperatures
given by the \Kepler\ Input Catalog \citep{2011AJ....142..112B},
so that we can present the rotational splitting as a function of
the stellar radius $R$  (Fig.~\ref{split_radius_masse}). The
evolutionary status of the stars was determined by
\cite{2012A&A...540A.143M}. Among the helium-burning stars, we
consider those with a mass greater than 1.8\,$M_\odot$ to belong
to the secondary clump. The uncertainties arising from the
effective temperatures and the imprecision of the scaling
relations do not impact the following analysis.

\subsection{Rotational splittings $\dnurot$}

Rotational splittings $\dnurot$ are shown as a function of the
large separation $\Dnu$ (Fig.~\ref{dnu_nurot}). We see that the
detection is difficult at low $\Dnu$, due to the limited frequency
resolution. This precludes an analysis of the core rotation in the
high-luminosity RGB stars, but allows us to measure rotation in
the clump stars.

We first consider the RGB stars, indicated by crosses in
Fig.~\ref{dnu_nurot}. We note that the rotational splitting
slightly decreases when $\Dnu$ decreases, that is, when the star
evolves on the RGB. For clump stars and secondary clump stars,
splittings are much smaller.
%Given that the rotational splitting is dominated by the \gmmode s, one would
%expect it will increase when the core contracts towards the tip of the RGB.
At this stage, ensemble asteroseismology indicates either that the
core rotation spins down, or that the splitting of \gmmode s is
not dominated by the core rotation. This last result is unlikely
since \gmmode\ splittings are significantly larger than \pmmode\
splittings (Fig.~\ref{exemple}), consistent with the observations
reported by \cite{2012Natur.481...55B} and
\cite{2012ApJ...756...19D} for four early RGB stars.

\subsection{Scaling relations}

We have examined how the rotational splittings evolve with the
stellar radius along the RGB (Fig.~\ref{split_radius_masse}). Only
RGB stars with a mass in the range [1.2, 1.5 $M_\odot$] were
considered, to avoid a bias from the fact that high-mass stars are
under-represented in the early stages of the RGB, whereas low-mass
giants are under-represented in the later stages. With 49 RGB
stars in this case, we find
\begin{equation}
 \dnurot \propto  R^{-0.5\pm0.3}\ \hbox{ (RGB).}
 \label{fit_RGB}
\end{equation}
In the first stages of the RGB, the splittings $\dnurot$ show a
slow decrease. Assuming the local conservation of angular
momentum, such a decrease seems in contradiction with the core
contraction: this has to be investigated.

The same exercise can be done for the clump stars. The fit,
conducted over a much broader range of mass
\citep{2012A&A...540A.143M}, gives
\begin{equation}
 \dnurot \propto  R^{-1.3\pm0.4}\ \hbox{ (clump) or }  \propto  R^{-1.4\pm0.4}\ \hbox{ (2nd clump).}
 \label{fit_clump}
\end{equation}
This indicates a different behaviour compared to RGB stars. We
first note that the slopes are independent of the stellar mass.
Closer to $-2$ than for RGB stars, they certainly relate the
influence of the stellar expansion. We also note that secondary
clump stars, which are more massive, show larger splittings than
clump stars.

\subsection{Intermediate conclusions}

From the analysis of the rotational splittings with the
stellar radius, we note the weak decrease of RGB stars. According
to the exponent of the fit reported by Eq.~\refeq{fit_RGB}, this
\gmmode\ splitting cannot be related to the surface rotation if it
evolves at constant local angular momentum. The large change in
the rotation evolution from the RGB to the clump can be related to
the expansion of the non-degenerate helium burning core
\citep{1971PASP...83..697I,2000ApJ...540..489S}. This increase of
the core radius is however limited and cannot explain the entire
observed decrease of the rotational splittings, so that we are
left with the most plausible conclusion that the strong decrease
of the rotational splitting is the signature of a significant
transfer of internal angular momentum from the inner to the outer
layers. This transfer should preferably occur at the tip of the
RGB, out of reach with current \Kepler\ observations due to a
limited frequency resolution. One could also imagine that the
rotational splittings are sensitive to different layers, depending
on the evolutionary status. This is investigated in the next
Section, where we aim to interpret the signification of the
observed splittings $\dnurot$.

%===================================================================
\section{Linking the rotational splittings to the core rotation\label{splitting-rotation}}

To go a step further, we intend to qualitatively link the observed
rotational splittings to the rotation inside the red giants.

\subsection{Linear rotational splittings and average rotation}

We assume that the rotation  is slow enough that a first-order
perturbation theory is sufficient to compute the rotational
splittings. This yields the following expression for rotational
splittings
\citep{1951ApJ...114..373L,1983SoPh...82..469C,1991sia..book..401C,2009LNP...765...45G,goupil}
\begin{equation}
\dnurotnl =  \int_0^1 \K\nl (x) ~\frac{\Omega(x)}{2\pi} \ \diff x,
\label{split}
\end{equation}
where $x=r/R$ is the normalized radius and $\Omega$ is the angular
rotation (in rad/s). The rotational kernel $\K\nl$ of the mode of
radial order $n$ and angular degree $\ell$ takes the form
\begin{equation}
 \K\nl =\frac{1}{I\nl}
  \Bigl[\xir^2+(\Lambda -1) \xih^2-2  \xir \xih\Bigr]\nl ~ \rho
  x^2\label{defK}
\end{equation}
where $I\nl$ is the mode inertia
\begin{equation}
I_{nl}= {\int_0^1 ~\Bigl[\xi_r^2+\Lambda\ \xi_h^2\Bigr]\nl~\rho
x^2 \diff x}.\label{defI}
\end{equation}
The quantities entering these equations are the fluid vertical and
horizontal displacement eigenfunctions, $\xir,\ \xih$
respectively; $\rho$ is the density, and $\Lambda= \ell(\ell+1)$.

The  mode inertia is much larger for \gmmode s that have a large
amplitude in the inner cavity than for \pmmode s. For a red giant,
several mixed modes exist in the frequency vicinity of each radial
mode. As a result, the variation with frequency of the dipole mode
inertia shows a regular variation with a periodicity roughly equal
to the large separation
\citep[see][]{2001MNRAS.328..601D,2009A&A...506...57D,jcd2011,
goupil}. This has been observationally confirmed
\citep{2012A&A...540A.143M}. The linear rotational splittings
(Eq.~\refeq{split}) are found to follow closely the same behavior
as the mode inertia  for the same reason. However a significant
amplification of the variation of the splittings with frequency
compared to that of mode inertia should exist when the rotation is
large in the central region \citep{goupil}.

\begin{figure}
\centering
\includegraphics[width=8.8cm]{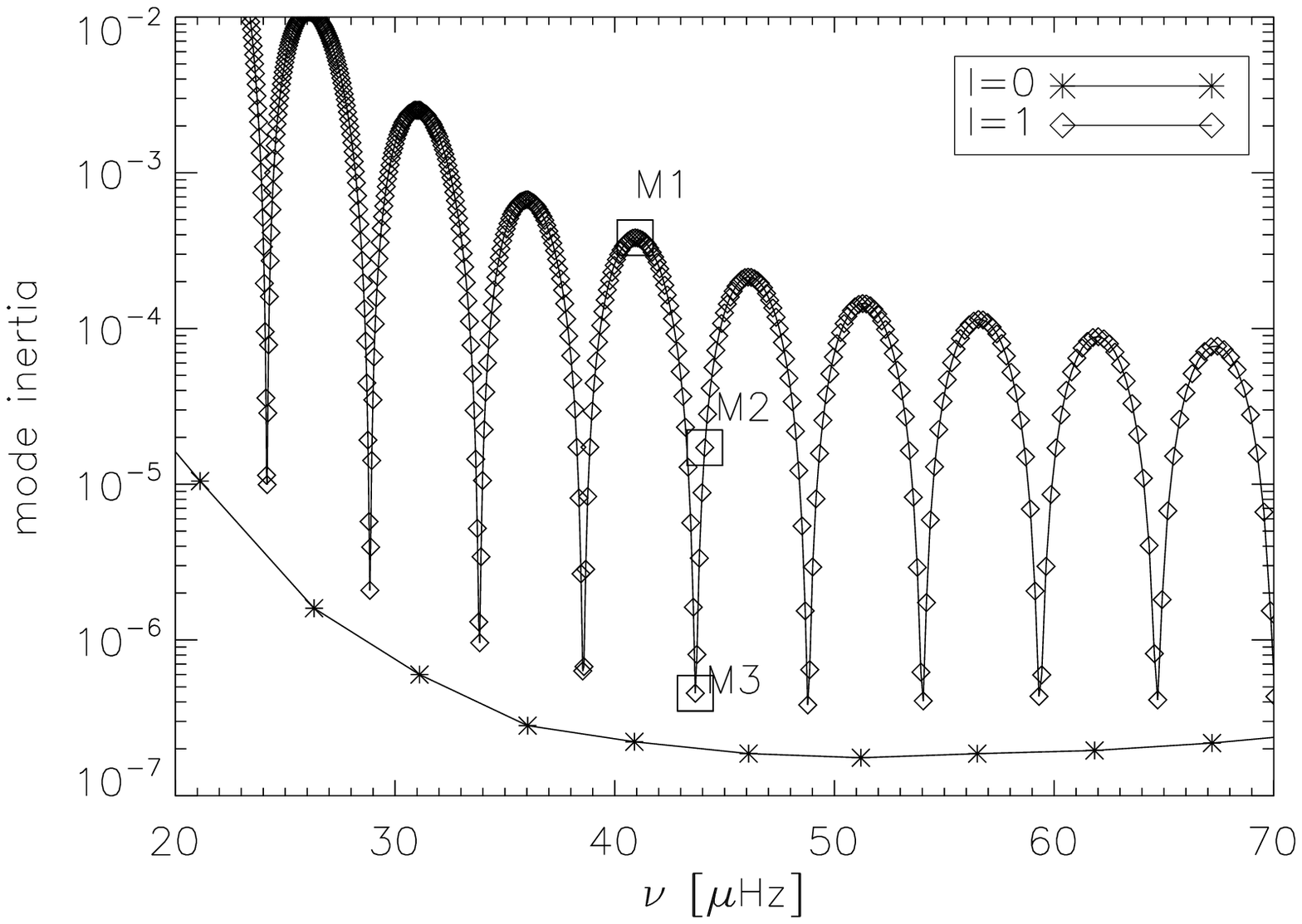}
\includegraphics[width=8.8cm]{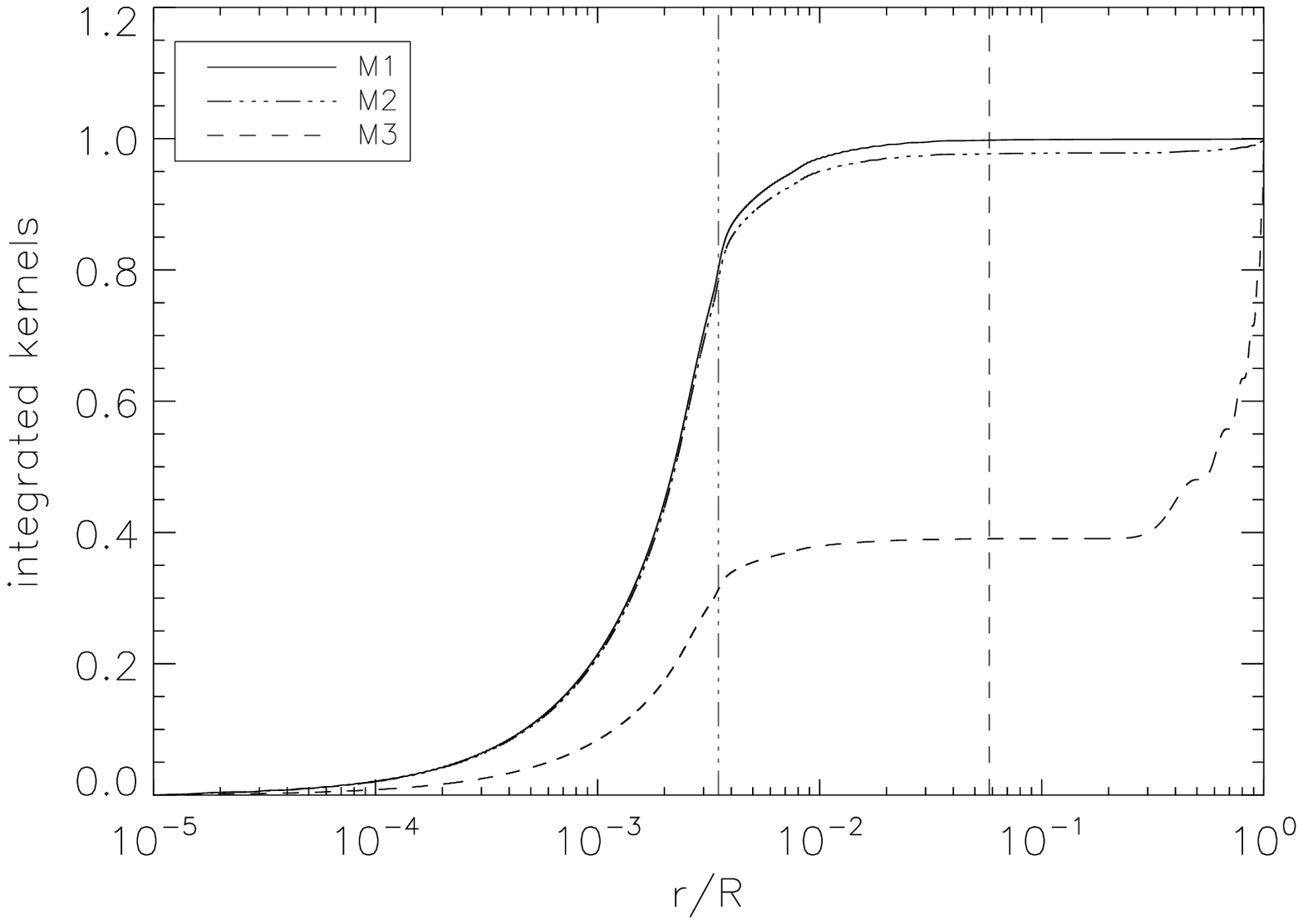}
\caption{{\sl Top:} mode inertia for radial and dipole modes of a
1\,$M_\odot$ RGB star at the bump. The evolution model has been
calculated with CESAM \citep{Morel1997} and the oscillation
frequencies were obtained with ADIPLS \citep{jcd2011}. {\sl
Bottom:} normalized integrated rotational kernels of three dipole
modes: $M_1$ and $M_2$ are \gmmode s, whereas $M_3$ is a \pmmode.
The vertical triple-dot-dashed line indicates the mean location of
the hydrogen-burning shell, and the vertical dashed line indicates
the base of the convective envelope.\label{kernels}}
\end{figure}

Because the red giants are characterized by an inner dense region
and an outer envelope, it is convenient to consider the rotational
splittings as the sum of two contributions
\begin{equation}
\dnurotnl =  {1\over 2 \pi} \Bigl(\OmegaK\corenl
 + \OmegaK\envnl \Bigr),
\label{E1}
\end{equation}
where $\OmegaK\corenl$ is the angular rotation, weighted by the
kernel, averaged over the central layers enclosed within a radius
$r\core$ and $\OmegaK\envnl$ the angular rotation averaged over
the layers above $\xcore=r\core/R$,
\begin{eqnarray}
 \OmegaK\corenl &\equiv & \int_0^{\xcore} \Omega(x)~ \K\nl(x) ~\diff x,\label{omegacrochet}\\
 \OmegaK\envnl &\equiv & \int_{\xcore}^1 \Omega(x)~\K\nl(x) ~\diff
 x.
\end{eqnarray}
The core boundary $\xcore$ must be understood here as the limit
where $\Omega(x)~ K\nl (x)$ no longer contributes to the integrant
in $\OmegaK\corenl$. Numerical calculations show that $x\core$
remains the same for all  modes \citep{joao}. Equation~\refeq{E1}
is then equivalent to
\begin{equation}
\dnurotnl = {\OmegaK\corenl\over 2\pi} ~\bigl(1 + \coefrotg
\bigr), \label{mini}
\end{equation}
where $\coefrotg$ can be written
\begin{equation}
\coefrotg = \left(\frac{K\envnl}{K\corenl}\right)~
\left(\frac{\Omegamoy\envnl}{\Omegamoy\corenl}\right),
\end{equation}
with the definitions
\begin{eqnarray}
K\corenl  & \equiv & \int_0^{\xcore} \K\nl(x) ~\diff x,  \\
K\envnl   & \equiv & \int_{\xcore}^1 \K\nl(x) ~\diff x,
\end{eqnarray}
and
\begin{eqnarray}
\Omegamoy\corenl &\equiv & \frac{\int_0^{\xcore} \Omega(x)
\ \K\nl (x) ~\diff x} {K\corenl}= {\OmegaK\corenl\over  K\corenl}, \label{rotmoys}\\
\Omegamoy\envnl  &\equiv & \frac{\int_{\xcore}^1 \Omega(x) \ \K\nl
(x) ~\diff x} {K\envnl}.
\end{eqnarray}
We then consider dipole \gmmode s that have the largest rotational
splittings; they coincide with  the largest inertia
\citep{2009A&A...506...57D}. These modes are furthest away from
the nominal pure pressure dipole modes and close to the radial
modes. The splittings associated with these modes do not vary with
frequency, i.e. from one radial mode to the other. Hence, we
retrieve the observed splittings (Eqs.~\refeq{splitt} and
\refeq{E1}):
\begin{equation}
\dnurot = \max \bigl( \dnurot{}_{,n,\ell=1} \bigr) = \max
\bigl(\dnusplit \bigr). \label{maxdnurot}
\end{equation}
As a consequence, we can drop the subscripts $n,\ell$ from now on.
One expects, for red giants, a more rapid rotation rate in the
inner layers hence $\Omegamoy\env/\Omegamoy\core< 1 $ and $\ll 1$
for very fast rotating cores \citep{goupil}. For the \gmmode s,
numerical calculations show that $K\env / K\core \ll 1$
(Fig.~\ref{kernels}), hence $\coefrotg \ll 1$. Thus, from
Eqs.~\refeq{mini} and \refeq{rotmoys}:
\begin{eqnarray}
 \dnurot &\simeq& {\OmegaK\core \over 2\pi}
 \simeq {\Omegamoy\core\over 2\pi}\  K\core.
\label{mini2}
\end{eqnarray}
For these modes, the displacement is essentially horizontal in the
core, therefore $\xir \ll \xih$. We also have $I \simeq I\core$,
so that, from Eqs.~\refeq{defK} and \refeq{defI}, one can derive
that the core kernel reduces to about $1/2$, in agreement with the
Ledoux coefficient of g modes \citep{1951ApJ...114..373L}.
Finally, we have
\begin{eqnarray}
\dnurot &\simeq& {1\over 2}{ \Omegamoy\core \over 2 \pi}.
\end{eqnarray}
Hence, rotational splittings of g-dominated modes provide a
measure of the rotation averaged over  the central region. The
averaged rotation roughly corresponds to the value of the rotation
at  the radius where the mode  amplitude of the horizontal
displacement $\xih$ is maximum. This happens away from the center,
in a core region where the rotation can have significantly
decreased compared to the central rotation. In that case, the
average rotation value gives a lower limit of the rotation of the
very deep layers. If the rotation happens to be nearly solid in
the central region, then the average rotation gives the rotation
of the nearly uniformly rotating core.

Keeping in mind these limits, $\dnurot$ can be considered as a
proxy of the mean rotation period of the core $\Trotcore$,
\emph{i.e.}
\begin{equation}\label{omegacor}
\Trotcore \equiv{ 2\pi \over \Omegamoy\core} \simeq {1 \over 2
\,\dnurot}
\end{equation}
for dipole mixed modes.

\subsection{Information from the kernels}

With the CESAM code for stellar evolution \citep{Morel1997} and
the ADIPLS code for adiabatic oscillations \citep{jcd2011}, we
estimate the kernels in red giant models at different evolutionary
states. Figure~\ref{kernels} shows the normalized integrated
rotational kernels $\int_0^r\! \K(r)\, \diff r / \int_0^R
\!\K(r)\, \diff r$ derived for a \pmmode\ and two \gmmode s in an
RGB star at the bump, with $\Dnu\simeq 5\,\mu$Hz. We verify that
the kernels in \gmmode s are dominated by the core, since the
normalized integrated kernels reach a value larger than 0.95 at
the core boundary.

These integrated kernels allow us to derive a refined estimate of
the core rotation. With a 2-layer model presented in
Appendix~\ref{appendix1}, a small correction to
Eq.~\refeq{omegacor} can be introduced by a factor $\coeff$
(Eq.~\refeq{facmesure})
\begin{equation}\label{corres_rot}
\Trotcore = {1 \over 2 \, \coeff\, \dnurot} .
\end{equation}
The correction factor $\coeff$, slightly larger than unity,
accounts for the propagation of the oscillation in the envelope.
This correction is intended to provide a more accurate result than
the proxy provided by Eq.~\refeq{omegacor}. Values of $\coeff$ can
be calculated for RGB stars at different evolution stages
(Appendix~\ref{appendix1}). They show that $\dnurot$ is less
dominated by the core rotation for early RGB stars compared to
more evolved stages.

As the value of $\coeff$ results from a balance between the
contributions of p and g modes, we can assume that its value
depends on the mean pressure and gravity radial orders. We
therefore propose a phenomenological proxy of the parameter
$\coeff$, justified by a simple model presented in
Appendix~\ref{appendix2}, as a function of global seismic
parameters. The analysis of the integrated kernels calculated for
red giant models at different evolutionary stages provides the fit
\begin{equation}\label{fit_eta_sismo}
   \coeff \simeq 1 + \gamma\ {\numax^2 \Tg \over \Dnu} ,
\end{equation}
where $\Tg$ is the period spacing of gravity modes derived from
the fit of the mixed modes and $\gamma \simeq 0.65$ (see Appendix
\ref{appendix2}). The model indicates that the coefficient
$\coeff$ decreases towards unity when a star evolves along the
RGB, due to the significant increase of the gravity radial orders
\citep{2012A&A...540A.143M}. The model, based on RGB stars, is
assumed to be valid also for clump stars, since it only relies on
global properties of the oscillation eigenfunction.

In all cases, the maximal splitting of \gmmode s is highly
dominated by the core rotation, and its measure is close to the
core rotation.  Furthermore, as found by
\cite{2012Natur.481...55B} and \cite{2012ApJ...756...19D}, we note
that there is no immediate link between the minimum splitting
$(1-\minir)\,\dnurot$ of \pmmode s (Eq.~\refeq{modulation}) and
the surface rotation. The minimum splitting measured for \pmmode s
is still strongly dominated by the core rotation. Extracting the
surface rotation, which is supposed to be small, requires a very
accurate description of the kernels, which is out of the scope of
this work dedicated to ensemble asteroseismology.

\begin{figure*}
\includegraphics[width=16cm]{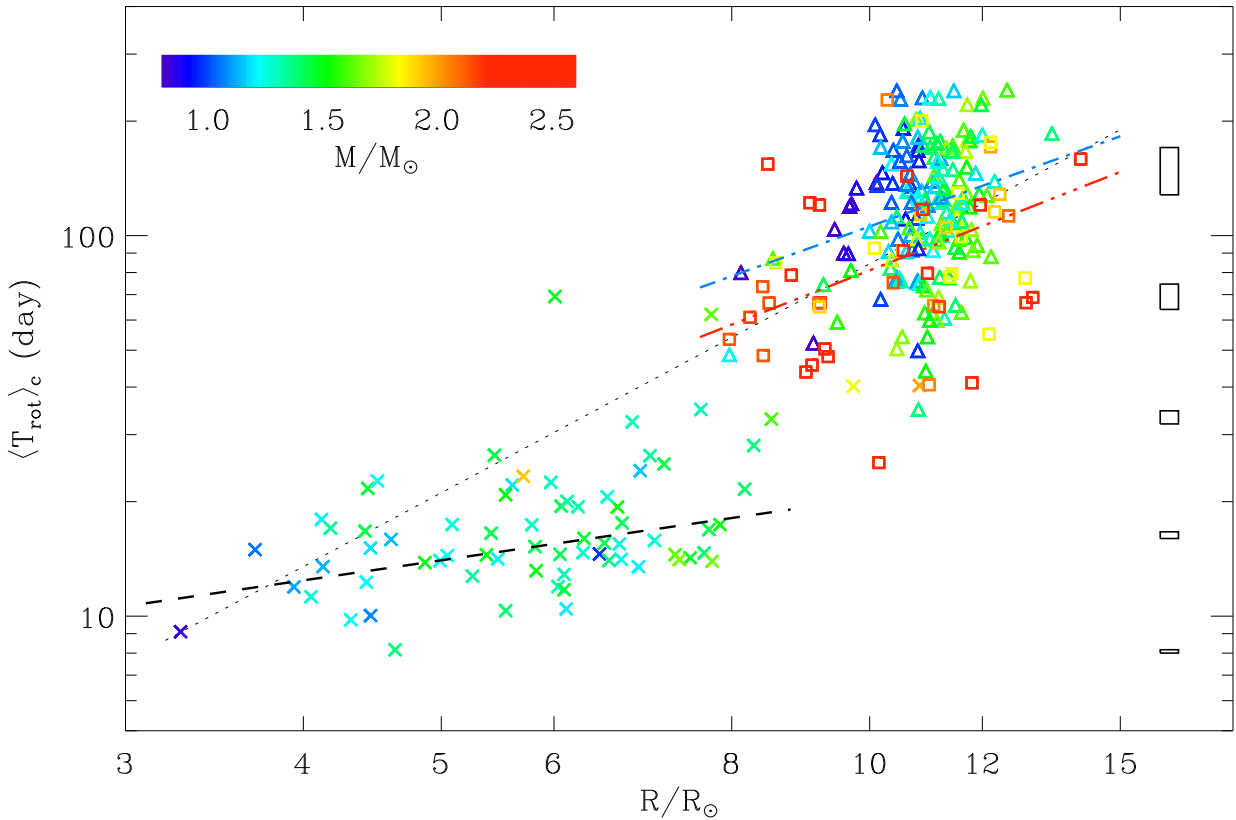}
\caption{Mean period of core rotation as a function of the
asteroseismic stellar radius, in log-log scale. Same symbols and
color code as in Fig.~\ref{dnu_nurot}. The dotted line indicates a
rotation period varying as $R^2$. The dashed (dot-dashed,
triple-dot-dashed) line indicates the fit of RGB (clump, secondary
clump) core rotation period. The rectangles in the right side
indicate the typical error boxes, as a function of the rotation
period. \label{rot_radius_masse}}
\end{figure*}

%=======================================================================
\section{Discussion\label{discussion}}

The relation between the maximum rotational splitting and the mean
core rotation period (Eq.~\refeq{omegacor}) allows us to revisit
the scaling relations established in Section \ref{scaling}. We
have to reiterate that Eq.~\refeq{corres_rot} is based on a strong
hypothesis, resulting from the linear relation between the
splitting and the core rotation rate. A high radial differential
rotation profile in the core, as shown by \cite{goupil} and
\cite{joao}, would invalidate the relation.

\subsection{Internal angular momentum transfer\label{transferRGB}}

The scaling relations in Eqs.~\refeq{fit_RGB} and
\refeq{fit_clump} can be written in terms of $\Trotcore$ rather
than $\dnurot$. We find, for RGB and clump stars, respectively:
\begin{eqnarray}
 \Trotcore \propto R^{0.7\pm0.3} & \ \hbox{ (RGB)}   \label{fit_RGB_T}\\
 \Trotcore \propto R^{1.4\pm0.4} & \ \hbox{ (clump)} \label{fit_clump_T}\\
 \Trotcore \propto R^{1.5\pm0.4} & \ \hbox{ (2nd clump)}. \label{fit_2clump_T}
\end{eqnarray}
We note that the absolute values of the exponents are similar to
the exponents found in Eqs.~\refeq{fit_RGB} and \refeq{fit_clump}.
This comes from the fact that the correction factor $\coeff$ of
Eq.~\refeq{fit_eta_sismo} is close to unity and does not show
important variation. As a consequence, the regime seen in the
rotational splitting translates into a similar regime of the mean
core rotation period (Fig.~\ref{rot_radius_masse}).

\subsubsection{On the RGB}

On the RGB, the spinning down of the core is moderate, much
smaller than a variation in $R^2$ expected in case of homologous
spinning down at constant total angular momentum. Angular momentum
is certainly transferred from the core to the envelope, in order
to spin down the core. However, a strong differential rotation
profile takes place when giants ascend the RGB
\citep{joao,goupil}.

\subsubsection{Clump stars}

The extrapolation of the fit reported by Eq.~\refeq{fit_RGB_T} to
a typical stellar radius at the red clump shows that cores of
clump stars are rotating six times slower. This slower rotation
can be partly explained by the core radius change occurring when
helium fusion ignition removes the degeneracy in the core. This
change, estimated to be less than 50\,\%
\citep{2000ApJ...540..489S}, can however not be responsible for an
increase of the mean core rotation period as large as a factor of
six. As a consequence, the slower rotation observed in clump stars
indicates that internal angular momentum has been transferred from
the rapidly rotating core to the slowly rotating envelope
(Fig.~\ref{rot_radius_masse}).

\subsubsection{Comparison with modeling}

The comparison with modeling reinforces this view \citep[Fig.~1
of][]{2000ApJ...540..489S}. Their evolution model assumes a local
conservation of angular momentum in radiative regions and
solid-body rotation in convective regions. It provides values for
the core rotation in a 0.8\,$M_\odot$ star of about 50\,days on
the main sequence, about 2\,days on the RGB at the position of
maximum convection zone depth in mass, and about 7 days in the
clump. This means that, even in a case where the initial rotation
on the main-sequence is slow (certainly much slower than the
main-sequence progenitors of the red giants studied here) and
where angular momentum is massively transferred in order to insure
that convective regions rotate rigidly, the predicted core
rotation periods are much smaller than observed. The expansion of
the convective envelope provides favorable conditions for internal
gravity waves to transfer internal momentum from the core to the
envelope to spin down the core rotation
\citep{1997A&A...322..320Z,2009A&A...506..811M}.
\cite{2008A&A...482..597T} have shown that the conditions are
favorable for these waves to operate at the end of the subgiant
branch and during the early-AGB phase. There is observational
evidence that the spinning down should have been boosted in the
upper RGB too.

The comparison of the core rotation evolution on the RGB and in
the clump shows that the angular momentum transfer is not enough
for erasing the differential rotation in clump stars. The line
representing an evolution of $\Trotcore$ with $R^2$ extrapolated
to typical main-sequence stellar radii gives a much more rapid
core rotation than the extrapolation from the RGB fit. This
indicates that the interior structure of a red-clump star has to
sustain, despite the spinning down of the core rotation, a
significant differential rotation. This conclusion, implicitly
based on the assumption of total angular momentum conservation, is
reinforced in case of total spinning down at the tip of the RGB.
However, the large similarities of the values of the core rotation
period observed in clump stars, together with an evolution of $
\Trotcore$ close to $R^2$ (Eq.~\refeq{fit_clump_T}), should imply
that a regime is found with a core rotation of clump stars much
more rapid than the envelope rotation but closely linked to it.

\subsection{Mass dependence}

We have calculated, for different mass ranges $[M_1, M_2]$, a mean
core rotation period defined by
\begin{equation}\label{mean_rot_m}
  \langle \Trotcore \rangle_{[M_1, M_2]} = {\sum_{M_1}^{M_2}\  \Trotcore \ R^{-r} \over \sum_{M_1}^{M_2} R^{-r}}
\end{equation}
where $r$ is the exponent given by Eqs.~\refeq{fit_RGB_T} or
\refeq{fit_clump_T}, depending on the evolutionary status.

This expression allows us to derive a mean value even for RGB
stars, in the mass range [1.2, 1.5\,$M_\odot$] where the RGB star
sample can be considered as unbiased. We do not detect any mass
dependence. The situation changes drastically for clump stars,
with a clear mass dependence: the mean value of $\Trotcore$ is
divided by a factor of about 1.7 from 1 to 2\,$M_\odot$. This
reinforces the view that angular momentum is certainly exchanged
in the upper RGB, since it may indicate a link between the amount
of specific angular momentum transferred and the evolution time:
high-mass red giants evolve more rapidly than lower mass stars,
loose less mass, and keep a more rapidly rotating core after the
tip.

%===================================================================
\section{Conclusion}

Rotational splittings were measured in about \nombreapprox\ red
giants observed during more than two years with \Kepler. As first
measured by \cite{2012Natur.481...55B} for three red giants in the
early stages of the RGB, a strong differential rotation is noted
for all these red giants.

We have first shown that the rotational splitting pattern,
modelled as a function of the mode frequency, is largely
independent of the stellar evolution. The empirical pattern found
by \cite{2012A&A...537A..30M} has been used and verified in a
large set of stars and has proven to be very efficient for
analysing the splittings. Independent of any modeling, we have
shown that the scaling relations observed for the maximum
rotational splittings in RGB stars may suggest that transfer of
angular momentum must occur in their interiors.

Then, assuming that the relation between the rotational splitting
and the rotation rate is linear, we have shown that the measured
splittings provide an estimate of a rotation period representative
of the mean core rotation. We observe that this period is larger
for clump stars compared to the RGB. This requires a transfer of
angular momentum in the star to spin down the core. Despite the
angular momentum loss expected at the tip of the RGB, the core
rotates more rapidly in clump stars than expected from an
evolution as the square of the radius. This indicates a strong
differential rotation in clump stars as well as in RGB stars. In
other words, the mechanism responsible for the redistribution of
angular momentum is efficient enough to spin down the mean core
rotation but with a time scale too long for reaching a solid
rotation.

The indirect estimate of the specific angular momentum shows that
massive red giants observed in the secondary clump have a
significantly higher specific angular momentum than in the main
red clump.

This ensemble asteroseismic analysis of rotation in red giants
will have to be extended to subgiants, since subgiants also show
mixed modes that give access to the inner rotation profile. As
\Kepler\ continues to observe, we will have access to longer
observation runs. This will provide more resolved observations of
the rotational splittings at low frequency, so we hope to measure
the mean core rotation on the upper part of the RGB and, if mixed
modes are also present, on the asymptotic giant branch. Our
findings provide strong motivation for further stellar modeling
including rotation.

%__________________________________________________________________
\begin{appendix}
\section{How rotational splittings are fitted\label{methode}}

\begin{figure}
\centering
\includegraphics[width=8.8cm]{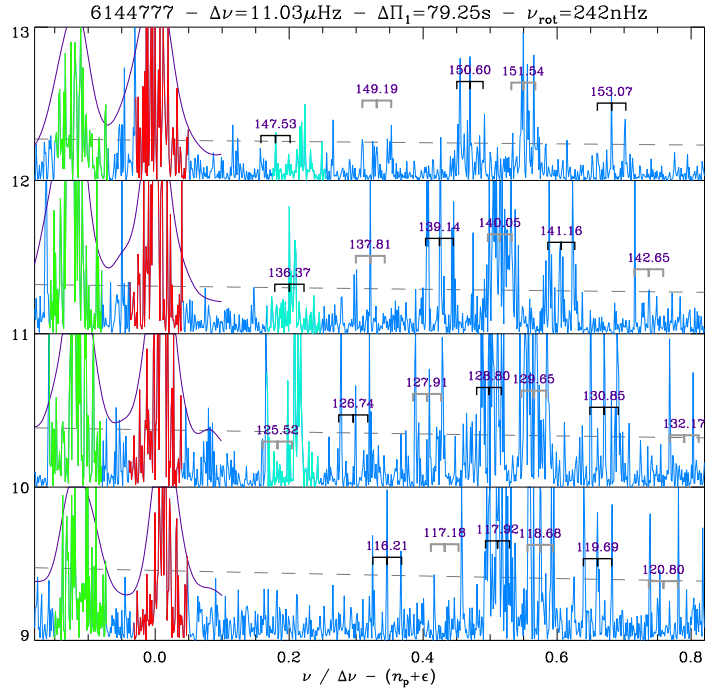}
\caption{Fit of rotational splittings, for the RGB star KIC
6144777, with an \'echelle diagram as a function of the reduced
frequency $\nu/\Dnu - (\np+\varepsilon)$. The radial orders are
indicated on the y-axis. Radial modes (highlighted in red) are
centered on 0, quadrupole modes (highlighted in green), near
$-0.12$ (with a radial order $\np-1$), and $\ell=3$ modes,
sometimes observed, (highlighted in hell blue) near 0.20.
Rotational splittings are identified with the frequency of the
$m=0$ component given by the asymptotic relation of mixed modes,
in $\mu$Hz. The fit is based on  peaks showing a height larger
than eight times the mean background value (grey dashed lines). In
order to enhance the appearance of the multiplets, highest peaks
have been truncated; to enhance the short-lived radial and
quadrupole modes, a smoothed spectrum is also shown, superimposed
on the corresponding peaks.\label{ex1}}
\end{figure}

\begin{figure}
\centering
\includegraphics[width=8.8cm]{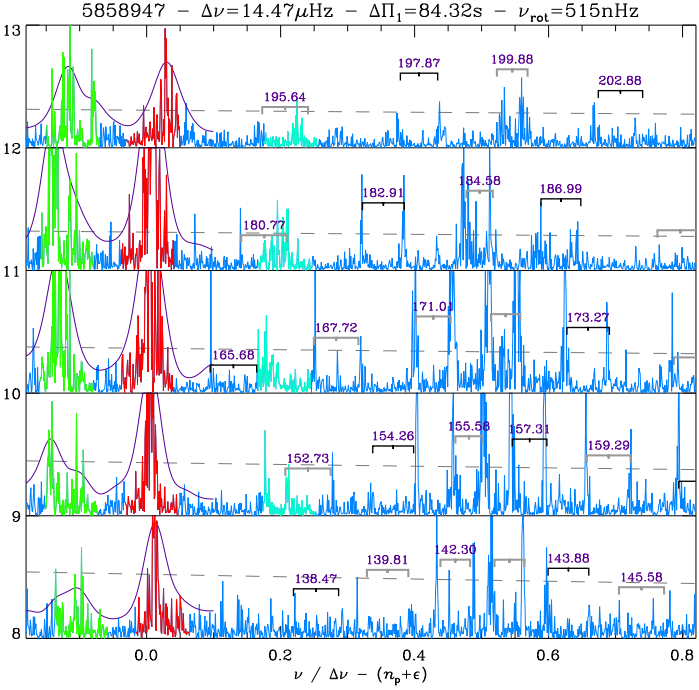}
\caption{Same as Fig. \ref{ex1}, for the RGB star 5858947. In such
a spectrum where the total splitting $2\dnurot$ is equal to half
the mixed-mode spacing at $\numax$, the fit allows to correctly
identify the multiplets. \label{ex1r}}
\end{figure}

\begin{figure}
\centering
\includegraphics[width=8.8cm]{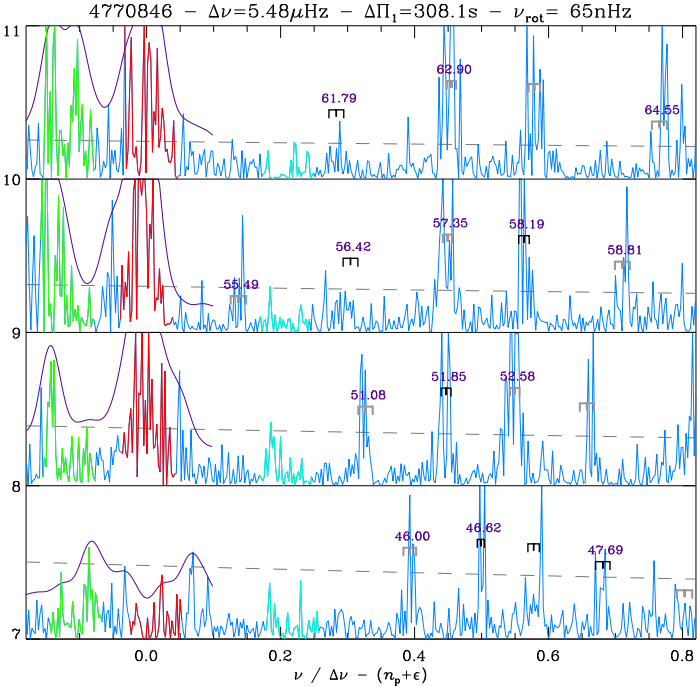}
\caption{Same as Fig. \ref{ex1}, for the clump star KIC 4770846.
The apparent low quality of the fit for \pmmode s at large
frequency is due to their short lifetimes
\citep{2011A&A...529A..84B}. \label{ex1c}}
\end{figure}

\begin{figure}
\centering
\includegraphics[width=8.8cm]{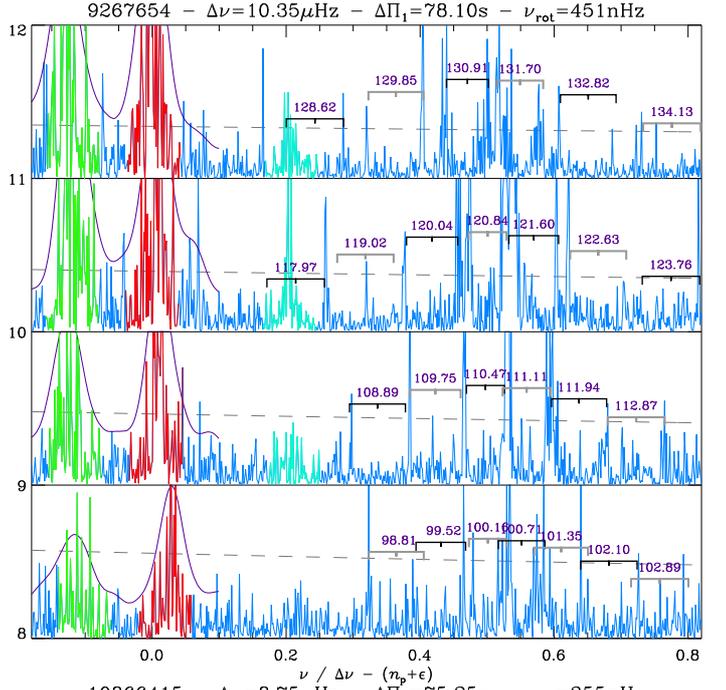}
\includegraphics[width=8.8cm]{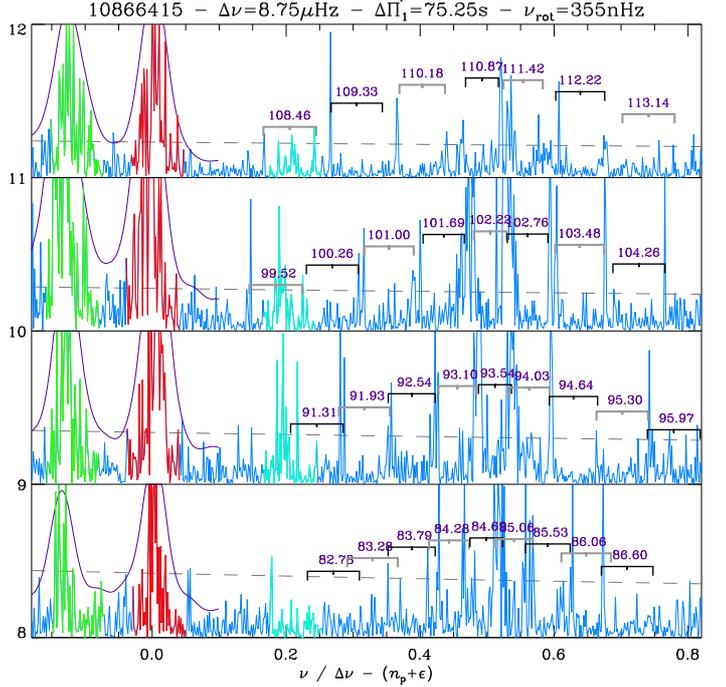}
\caption{Same as Fig. \ref{ex1}, for the RGB stars KIC 9267654 and
KIC 10866415, where the total splitting $2\dnurot$ is nearly equal
to the mixed-mode spacing at $\numax$. Apparent narrow multiplets
are artifacts due to close combinations between components of
different mixed-modes radial orders. \label{ex2}}
\end{figure}

\begin{figure}
\centering
\includegraphics[width=8.8cm]{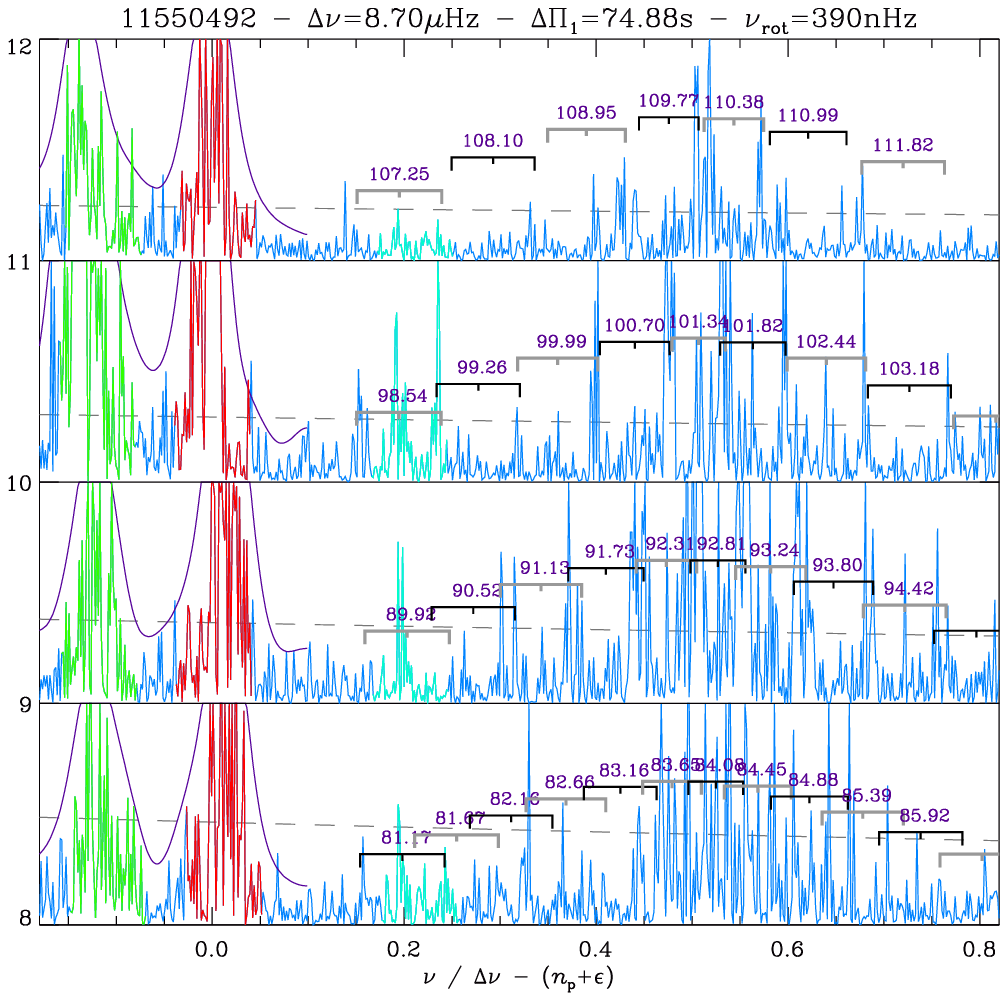}
\caption{Same as Fig. \ref{ex2}, for the RGB star KIC 11550492.
The non-negligible amplitudes of the $m=0$ components complicate
the analysis. \label{ex3}}
\end{figure}

\subsection{Large separation and gravity mode spacing}

The first step for identifying the red giant oscillation spectrum
is, as for all stars showing solar-like oscillations, the correct
identification of the radial mode pattern, in order to locate
precisely the location of the theoretical pure dipole pressure
modes. The fit of the radial modes depends mainly on the accurate
determination of the large separation. According to the universal
red giant oscillation pattern \citep{   2011A&A...525L...9M}, the
surface offset and the curvature of the ridge are functions of the
large separation. In practice, small residuals due to glitches
\citep{2010A&A...520L...6M} can induce a frequency offset of
about, typically, $\Dnu / 50$. Thus, a second free parameter,
simply a frequency offset, or equivalently an offset of
$\varepsilon$ less than 0.02 (Eq.~\refeq{tassoulp}), is useful for
providing the best fit of the radial ridge. The location of the
dipole ridge with respect to the radial ridge is given by the
small separation $\d01$ (Eq.~\refeq{tassoulp}), which is a
function of the large separation \citep{2011A&A...525L...9M}.

The fit of the mixed-mode pattern is based on two free parameters:
the period spacing $\Tg$ and the coupling constant $q$, as defined
by Eq.~9 of \citep{2012A&A...540A.143M}, which closely follows the
formalism of mixed modes given by \cite{1989nos..book.....U}. In
order to determine $\Tg$ on the RGB, it is worthwhile to consider
that this period is a function of the large separation. For the
low-mass stars of the RGB with a degenerate helium core, a
convenient proxy is given by the polynomial development
\begin{equation}\label{Dnu-Dpi1}
    \Tg = 62.5 + 1.40\, \Dnu + 0.081\, \Dnu^2
\end{equation}
with $\Dnu$ in $\mu$Hz and $\Tg$ in s, according to Fig.~3 of
\cite{2012A&A...540A.143M}. When the rotational splitting is
larger than half the mixed mode spacing at $\numax$, this step
cannot be done independent of the next one.

\begin{figure*}
\centering
\includegraphics[width=7cm]{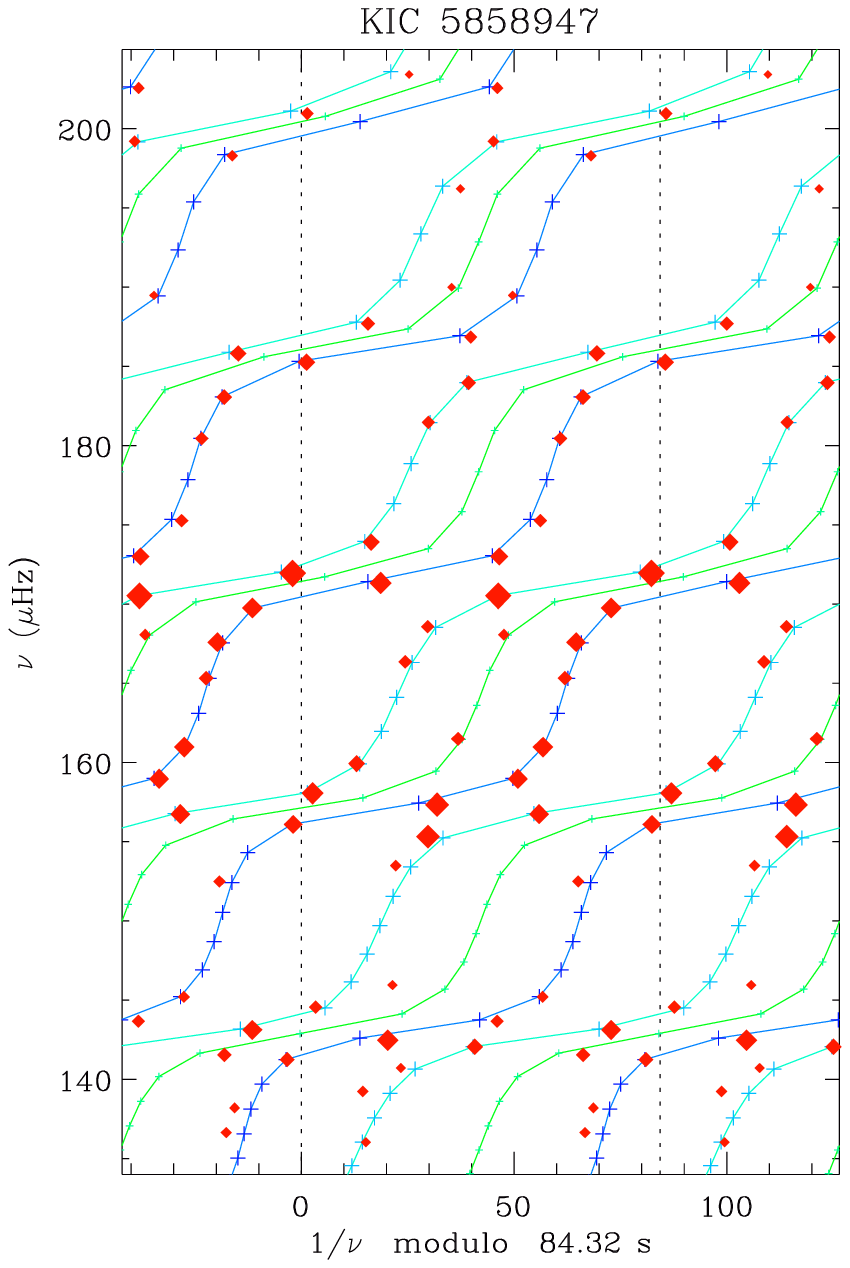}
\includegraphics[width=7cm]{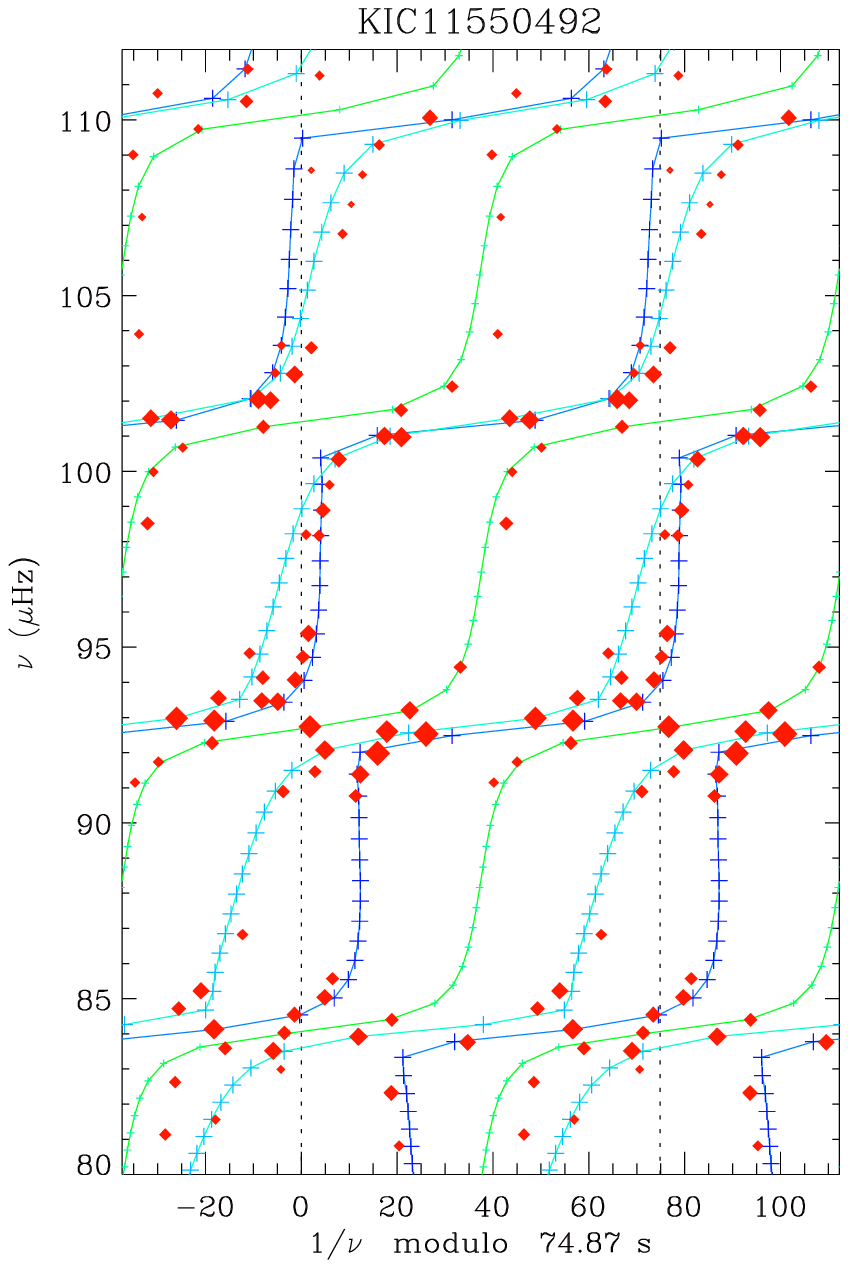}
\caption{Gravity \'echelle diagrams of the two RGB stars KIC
5858947 and 11550492. The x-axis is the period $1/\nu$ modulo the
gravity spacing $\Tg$; for clarity, the range has been extended
from $-0.5$ to 1.5 $\Tg$. The size of the selected observed mixed
modes (red diamonds) indicates their height. Plusses give the
expected location of the mixed modes, with $m=-1$ in light blue,
$m=0$ in green and $m=+1$ in dark blue.  \label{degech}}
\end{figure*}

\subsection{Rotational splittings}

Great care must be taken to disentangle the splittings from the
mixed mode spacings. Three major cases have to be considered for
fitting the rotational splittings.

- If splittings are small and almost uniform with frequency,
except the modulation depicted by $\Frot$ (Eq.~\ref{modulation}),
then the estimate is straightforward. The unknown stellar
inclination can be derived from the mode visibility, which depends
on the azimuthal order $m$. According to the probability of having
an inclination $i$ proportional to $\sin i$, in most cases
doublets with $m=\pm 1$ are observed. Note that, even if the
components $m=-1$ and $+1$ have the same visibility, they may in
practice present different heights, due to the stochastic
excitation of the modes. Such splittings smaller than the
mixed-mode spacings are seen in the lower stages of the RGB and in
the clump (Fig.~\ref{ex1}).

- If \emph{apparent} splittings at $\numax$ seem to increase with
increasing frequency, then the most plausible solution is that
$\dnurot$ is close to half the mixed-mode spacing at $\numax$.
These apparent splittings result in fact from a mixing of the
splittings embedded with the spacings. Such a situation occurs
when the apparent splittings are composed of the $m=\pm 1$
component of the mixed mode order $\nm$ and of the $m=\mp 1$
component of the adjacent orders $\nm\pm 1$. The true splittings,
significantly larger than the apparent splittings, are almost
uniform for \gmmode s. This uniformity is used for iterating the
solution. Such cases occur most often for RGB stars with $\Dnu$ in
the range [9 -- 12\,$\mu$Hz] (Fig.~\ref{ex2}).

- If \emph{apparent} splittings seem very irregular, then the most
plausible solution is that $\dnurot$ is much larger than half the
mixed-mode spacing at $\numax$. In fact, the apparent splittings
are complex structures resulting from a mixing of components of
two or three different mixed-mode orders. A careful visual
inspection is necessary to disentangle them. The mixed-mode
asymptotic expression and the empirical expression of the
rotational splitting are accurate enough for resolving complex
cases that occur for RGB stars with $\Dnu \le 9\,\mu$Hz
(Fig.~\ref{ex3}).

We have used gravity \'echelle diagrams to represent the mixed
modes \citep{2011Natur.471..608B,2012A&A...540A.143M}. Due to the
complexity of the features caused by embedded splittings and mixed
modes spacings, the \'echelle diagrams cannot be used to identify
the rotational splittings, but are useful for improving the
accuracy of the fit. In the examples shown (Fig.~\ref{degech}), a
10-s shift between the periods of the observed and modeled peaks
correspond to an accuracy in frequency of about $\Dnu/100$.

\subsection{A dipole mode forest?}

The complete fit of the rotational splittings is based on three
parameters: the maximum splitting $\dnurot$ and the two parameters
$\minir$ and $\beta$ entering the definition of $\Frot$. The best
fit is provided by correlating the observed multiplets with
synthetic multiplets.

Since the parameters $\minir$ and $\beta$ are found to vary in
narrow ranges, the solution for inferring $\dnurot$ (and
simultaneously $\Tg$ on the RGB with low $\Dnu$) is based on
considering them as constants. As a result, five free parameters
are enough for fitting the whole red giant oscillation spectrum.
Variation of $\minir$ and $\beta$ allows a better fit. The stellar
inclination can be derived from the ratio of the visibility of the
$m=\pm1$ components compared to the central component.

In a typical spectrum, more than 30 mixed-mode orders,
representing about 60 to 120 individual modes with a height larger
than eight times the background are simultaneously fitted. The
typical accuracy of the fit, of about $\Dnu/200$ or better, is
enough for avoiding any confusion in almost all cases, except for
the most evolved RGB stars.

Finally, with the identification of the mixed mode spacings and of
the rotational splittings, the dipole mode forest becomes a
well-organized garden \emph{\`a la fran\c caise}.

\section{Two-layer model\label{layers}}

\subsection{Core and surface contributions\label{appendix1}}

In order to estimate the contribution of the core and surface
rotation, we simplify the stellar stratification to a 2-layer
model. We denote by $\rotc$ and $\rots$ the rotational frequency
of the core and at the surface, respectively, and $\rotg/2$ and
$\rotp$ the measured splitting on g and p modes, respectively. The
factor 1/2 in $\rotg/2$ accounts for the Ledoux coefficient. The
contributions of the surface and of the core are written
\begin{equation}\label{som}
\left\{\begin{array}{rcclccl}
  \rotg &=& \xg  & \rotc   &+& (1-\xg) &\rots \\
  \rotp &=& \xp  & \rotc   &+& (1-\xp) &\rots \\
\end{array}\right. .
\end{equation}
The coefficient $\xp$ and $\xg$ are derived from the rotational
kernels. From the solution
\begin{equation}\label{sol}
    \rotc = {1-\xp\over \xg-\xp} \rotg + {\xg-1\over \xg-\xp} \rotp
\end{equation}
and from the observation of the splitting of \pmmode s indicating
$\rotp \simeq \rotg / 4$ (a factor of about 1/2 comes from
$1-\minir$ in Eq.~\refeq{modulation}, an another factor of 1/2
comes from the Ledoux coefficient), one derives that the measure
of $\rotg$ is an indicator of the core rotation
\begin{equation}\label{facmesure}
\rotc = \coeff\ \rotg .
\end{equation}
For an RGB star at the bump with $\Dnu=5\,\mu$Hz, the values $\xp$
and $\xg$ derived from the kernels give $\coeff=1.06\pm0.04$, very
close to unity. A less evolved star, as considered by
\cite{2012Natur.481...55B}, whose mixed modes correspond to much
smaller radial gravity orders, has $\coeff=1.45^{+0.30}_{-0.15}$.
\cite{2012ApJ...756...19D} derived a similar result for a giant
with $\Dnu \simeq 29\,\mu$Hz at the bottom of the RGB. This shows
that $\dnurot$ is less dominated by the core rotation for early
RGB stars. One also derives that, in all cases, the surface
rotation $\rots$ is small, and that measuring it precisely from
the \gmmode -splitting is not possible.

\subsection{Link to the eigenfunction properties\label{appendix2}}

The value of the coefficients $\xg$ and $\xp$ introduced in
Eq.~\refeq{som} can be approximated by the expression of the
rotational splitting (Eqs.~\refeq{split} and \refeq{defK}).
Basically, the integration of the wave function has a contribution
varying as the number of nodes in the core and in the envelope,
respectively. As a consequence, $\xg \propto \ng$ and $\xp \propto
\np$. In order to more precisely account for the complex form of
the wave function, we suppose:
\begin{equation}\label{somm}
\left\{\begin{array}{rcccl}
  \xg &=& \gg\ng &/& (\gg\ng + \gp\np)   \\
  \xp &=& \gp\np &/& (\gg\ng + \gp\np)   \\
\end{array}\right.
\end{equation}
with $\gg<0$ to account for the negative value of $\ng$. The
validity of this development implicitly assumes that $\gp$ and
$|\gg|$ are constant not so far from unity. Hence, neglecting
$\rots$ in Eq.~\refeq{som}, we derive:
\begin{equation}\label{fin_approx}
\rotc \simeq \left[ 1 + {\gp\over\gg} {\np\over\ng}\right] \rotg
\simeq \left[ 1 - \gamma {\np\over\ng}\right] \rotg .
\end{equation}
The radial order $\np$ and $\ng$ have to be estimated at the
frequency $\numax$ where the oscillation amplitude is maximum.
Then, we can derive that $\coeff=\rotc / \rotg$ is related to the
global seismic parameters $\Dnu$ and $\Tg$, where $\Tg$ is period
spacing of gravity modes:
\begin{equation}\label{fit_eta_sismo_app}
   \coeff \simeq 1 + \gamma\ {\numax^2 \Tg \over \Dnu}
\end{equation}
The fit of the integrated kernels calculated at different
evolutionary stages gives $\gamma \simeq 0.65$.  This
phenomenological result based on a simple two-layer model is to be
considered as a proxy only.

\end{appendix}
%______________________________________________________________
\begin{acknowledgements}

Funding for this Discovery mission is provided by NASA's Science
Mission Directorate. This work partially used data analyzed under
the NASA grant NNX12AE17G. NCAR is supported by the National
Science Foundation. PGB has received funding from the European
Research Council under the European Community's Seventh Framework
Programme (FP7/2007-2013)/ERC grant agreement n$\circ$ PROSPERITY.
SH acknowledges financial support from the Netherlands
Organisation for Scientific research (NWO). YE acknowledges
financial support from the UK STFC. RAG acknowledges the support
of the European Community's Seventh Framework Program
(FP7/2007-2013) under grant agreement no. 269194 (IRSES/ASK).

\end{acknowledgements}

\bibliographystyle{aa} % style aa.bst
\bibliography{biblio_rotation}

\end{document}